\documentclass[preprint]{JASA}

\usepackage{graphicx}
\usepackage{acronym}
\usepackage{xspace}

\usepackage{amsfonts}
\usepackage{float}

\let\linenumbers\nolinenumbers\nolinenumbers  

\begin{document}

\title[Compact representation of echosounder time series]{Compact representation of temporal processes in echosounder time series via matrix decomposition}

\author{Wu-Jung Lee}
\email{wjlee@apl.washington.edu}
\affiliation{Applied Physics Laboratory, University of Washington, Seattle, WA 98105, USA.}
 
\author{Valentina Staneva}			
\affiliation{eScience Institute, University of Washington, Seattle, WA 98105, USA.}

\preprint{}		


\begin{abstract}
\hspace{1cm} This paper is part of a special issue on Machine Learning in Acoustics.

\noindent \textbf{Abstract}

The recent explosion in the availability of echosounder data from diverse ocean platforms has created unprecedented opportunities to observe the marine ecosystems at broad scales. However, the critical lack of methods capable of automatically discovering and summarizing prominent spatio-temporal echogram structures has limited the effective and wider use of these rich datasets. To address this challenge, we develop a data-driven methodology based on matrix decomposition that builds compact representation of long-term echosounder time series using intrinsic features in the data. In a two-stage approach, we first remove noisy outliers from the data by Principal Component Pursuit, then employ a temporally smooth Nonnegative Matrix Factorization to automatically discover a small number of distinct daily echogram patterns, whose time-varying linear combination (activation) reconstructs the dominant echogram structures. This low-rank representation provides biological information that is more tractable and interpretable than the original data, and is suitable for visualization and systematic analysis with other ocean variables. Unlike existing methods that rely on fixed, handcrafted rules, our unsupervised machine learning approach is well-suited for extracting information from data collected from unfamiliar or rapidly changing ecosystems. This work forms the basis for constructing robust time series analytics for large-scale, acoustics-based biological observation in the ocean.

\end{abstract}

\maketitle

\raggedbottom

\section{Introduction}
\label{sect:introduction}

Sound is extensively used to study life in the ocean \cite{Medwin1998}. Compared to net-based sampling, which can only be conducted at discrete times and locations, echosounders boast the capability to ``connect the dots'' by delivering continuous active acoustic observation across time and space. This advantage has made them an indispensable tool in modern  ecological and fisheries studies, in particular for collecting information about mid-trophic level organisms that are otherwise difficult to observe effectively at large scale \cite{Handegard2013, Benoit-Bird2016}.

\begin{figure}[!htb]
	\centering
	\includegraphics[width=0.5\textwidth] {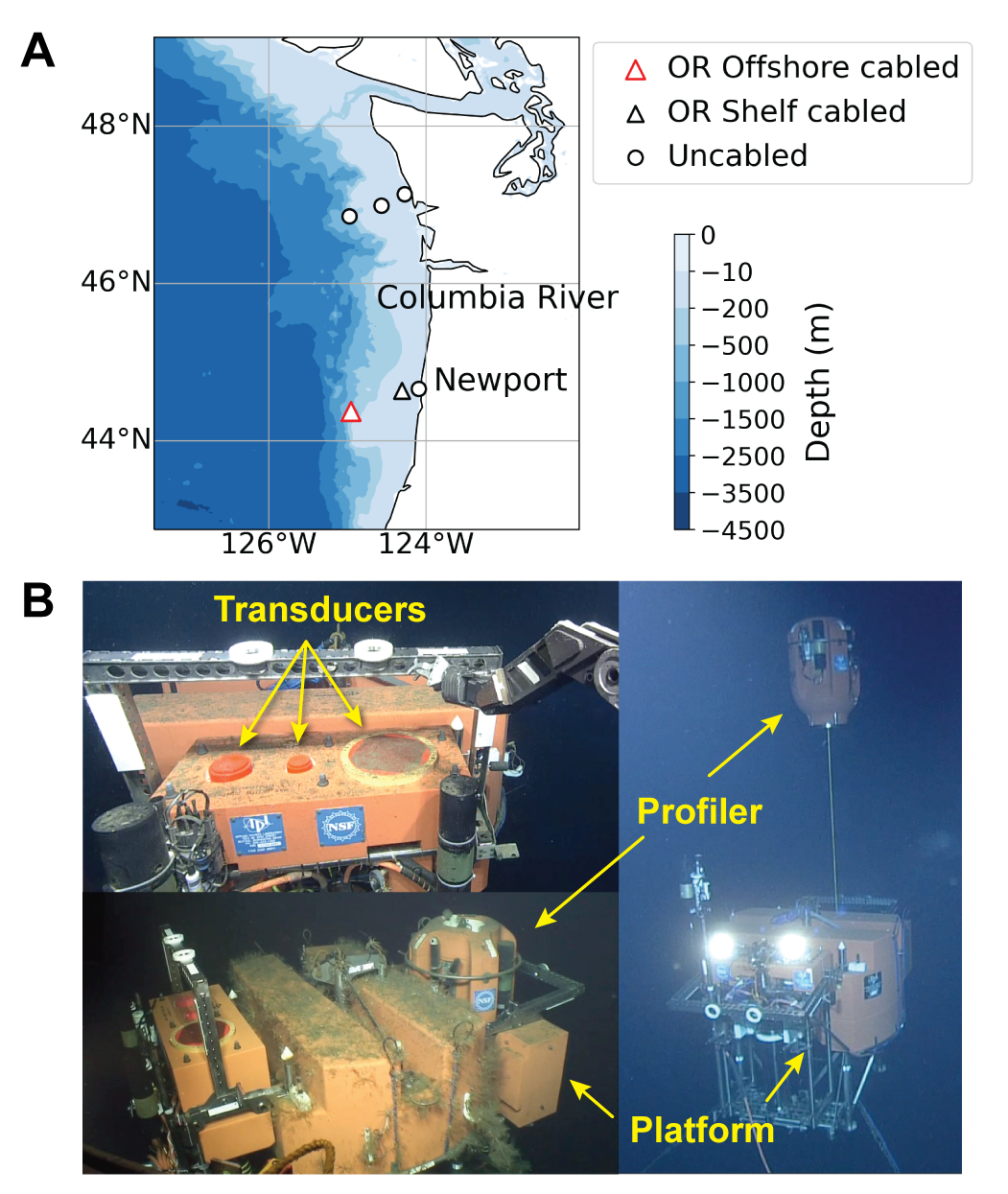}
	\caption{(A) Data used in this work were collected by a three-frequency echosounder installed on a moored underwater platform located at the Oregon Offshore site of the Ocean Observatories Initiative (OOI) Coastal Endurance Array (red triangle). The symbols indicate the locations of all OOI echosounders installed along the coasts of Oregon and Washington. (B) The transducers are integrated into the mooring platform (from left to right: 120, 200, and 38 kHz). The platform also hosts a profiler that traverses the water column above the echosounder. (Image credit: University of Washington, Ocean Observatories Initiative---NSF, Woods Hole Oceanographic Institution.)}
	\label{fig:data_source}
	\vspace{0.5cm}
\end{figure}

Technological advancements have resulted in a deluge of echosounder data in the past decade, thanks to the rapid and successful integration of autonomous echosounders with a wide variety of ocean observing platforms, including moorings and autonomous surface and underwater vehicles \cite{Greene2014, Suberg2014, Moline2015, Dunlop2018, DeRobertis2018_mooring}. The significantly broader spatial and temporal coverage of the datasets provides an unprecedented opportunity to study the response of the marine ecosystems to climate variability at scales never before possible. An example is the network of six moored echosounders maintained by the Ocean Observatories Initiative (OOI) along the coast of Northeast Pacific, which have been delivering data continuously since 2015 (Fig.~\ref{fig:data_source}A). Another example is the experimental fisheries survey carried out by a fleet of autonomous surface vehicles sampling the west coast of the U.S. and Canada simultaneously in the summer of 2018-2019 \cite{Chu2019}. However, the massive volume and complexity of these new data have overwhelmed the conventional echosounder data analysis pipelines that rely heavily on manual processing, thwarting rapid progress in marine ecological research.

A critical need in the analysis of large echosounder datasets is the automatic extraction of biological information, such as organism identity, biomass, and high-level spatio-temporal distribution patterns of organisms, without excessive human intervention. Currently, the majority of echo data processing procedures require visual inspection and labeling of echograms (image formed by echo returns, e.g., Fig.~\ref{fig:echogram_raw_pcp}A) by human experts \cite{hake_survey_protocol_2012}, prompting not only scalability issues with increasing data volume but also subjective bias and reproducibility challenges. 

Many echo classification methods were proposed to automate the procedure of discerning the taxonomic identity of acoustically observed biological aggregations. A key feature commonly exploited is the relative strength of echo across sonar frequency, which varies significantly as a function of the size, shape, orientation, and material properties of marine animals \cite{Medwin1998, Benoit-Bird2016}. These include classification rules derived using empirical echo measurements and physics-based scattering models, such as the popular frequency-differencing and related Bayesian inference methods \cite[e.g.,][]{Jech2006, DeRobertis2010_multifreq, Korneliussen2016}, as well as supervised and semi-supervised machine learning methods, such as random forests and neural networks \cite[e.g.,][]{Haralabous1996, Woillez2012, Fallon2016, Brautaset2020}. 
However, these classification methods can be ill-fitted for large-scale echosounder data collected on autonomous platforms, due to challenges associated with collecting concurrent biological ground truth and calibration information at a scale comparable to that of the acoustic data. For example, extensive net- or optics-based biological taxa identification samples are critical in constructing empirical reference or training datasets for scatterer classification, but the cost and logistical complexity involved in collecting them in parallel with autonomous echosounder sampling is often prohibitive, especially in geographically remote areas or deep water. The validity and applicability of any derived classification rules further hinge on the assumed invariance of the combination of measurement systems and the biological community, both of which can vary significantly across time and space. In addition, accurate echosounder calibration---the mapping between the received voltage signal and the actual echo strength in physical units \cite{Demer2015, Haris2018}---imposes a stringent requirement for adequate application of these methods, as calibration needs to be conducted separately for each frequency under a wide range of environmental conditions (e.g., at depth) under which acoustic data are collected. 

Another suite of methods were developed with a goal of capturing specific biological activity patterns observed in the echogram. These methods seek to capture local echogram features through summary statistics derived from a vertical echogram slice or across a few immediately adjacent slices. The features include integrated echo energy (as a proxy for abundance), the vertical spread of animal distribution, timing and speed of animals engaged in vertical migration, and the number and depth of strong scattering layers, etc. \cite[e.g.,][]{Woillez2007, Urmy2012, Cade2014, Proud2015, Klevjer2016}. While useful in summarizing echogram variations, these statistics are handcrafted and not adaptive, risking loss of information or poor characterization when the pre-determined echogram features mismatch those of the actual data. The summary statistics also lack the capability to automatically extract and describe high-level echogram features since the computation involves only temporally and spatially localized measurements in large datasets. A recent study circumvented some of these difficulties by employing a time-augmented extension of the empirical orthogonal function method to examine the anomaly of zooplankton diel vertical migration behavior with respect to monthly means \cite{Parra2019}. The anomaly patterns were subsequently linked to factors such as ocean currents, lunar variability and the composition of the local biological community.

Rather than focusing on specific organisms or localized echogram features, the overarching goal of our work is to develop a methodology that can automatically build a comprehensive representation of high-level spatio-temporal structures intrinsic to the echosounder data, such as temporally transient appearance of scattering layers or animal aggregations engaged in different vertical movement behaviors. These structures are the most prominent and immediate features human observers attend to in visual analysis of large echograms, and are key to forming and testing hypotheses pertinent to organism responses to environmental disturbance of natural or anthropogenic sources. 

In this paper, we show that matrix decompositions are powerful techniques capable of achieving this goal in an unsupervised manner with minimal assumption and prior knowledge from the observed region. We show how different decomposition formulations are suitable for exploiting and extracting different structures in the data: 1) Principal Component Pursuit (PCP) is a robust version of Principal Component Analysis capable of exploiting the latent regularities in long-term echograms to automatically remove spurious and noisy echo outliers while retaining the high-level spatial and temporal patterns present in the original time series; 2) temporally smooth Nonnegative Matrix Factorization (tsNMF) is a decomposition with nonnegativity and temporal smoothness constraints that successfully extracts distinct daily echogram patterns with time-varying activation sequences, providing a compact representation of temporal processes embedded in the data. The utility of these methods are demonstrated using a three-frequency echosounder time series spanning 62 days. While no specific constraints are incorporated into the decomposition formulation in this paper to account for multi-frequency echo features, we discuss preliminary results from a recent work \cite{Lee2019_tensor} in which we investigated how these features may be exploited through nonnegative tensor decomposition, a multi-way extension of nonnegative matrix factorization, as well as other avenues for future development.

Our contribution builds on the classic concept of dimensionality reduction to find structures directly from the data and use them for representation and interpretation. In contrast to previous methods that depend strongly on past experience and observer assumptions, the unsupervised data-driven approach we employ is adaptive to intrinsic properties of the data. Such a capability is pivotal in extracting and synthesizing information from the overwhelming volumes of echosounder data collected from not only unfamiliar ecosystems but also those undergoing rapid changes in the changing climate.

\section{Materials and Methods}

\subsection{Acoustic and environmental data} 
\label{sect:echo_data}
We use 62 days of acoustic data (August 17--October 17, 2015) from an upward-looking echosounder (Simrad EK60) mounted on a moored underwater platform at a depth of approximately 200 m. This echosounder is part of the Oregon Offshore Cabled Shallow Profiler Mooring (\citeauthor{OOI_CE04OSPS}) of the OOI Coastal Endurance Array (Fig.~\ref{fig:data_source}A), on which data are collected continuously and become immediately available over the Internet. The full water depth at this site is 588 m.

\begin{figure}[!htb]
	\centering
	\includegraphics[width=1\textwidth]{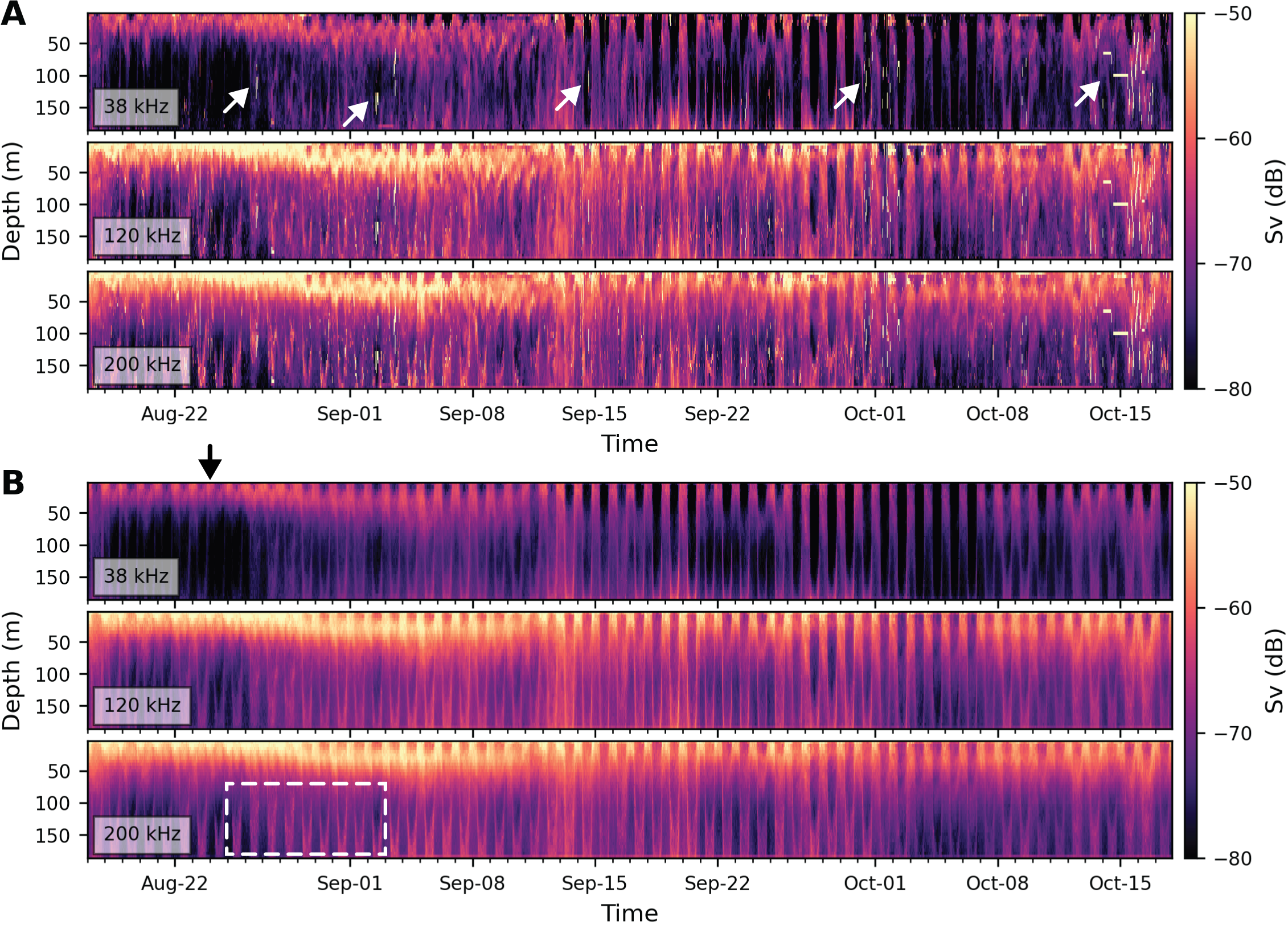}
	\caption{(A) Multi-frequency mean volume backscattering strength (MVBS) time series collected by the echosounder. The white arrows indicate echoes from the water column profiler. (B) The low-rank component resulting from Principal Component Pursuit (PCP) analysis of the MVBS time series. The black arrow indicates the time period during which the mid to upper water column was relatively ``empty'' at 38 kHz. The dashed box highlights the gradual depth variation of the diel vertical migration pattern.}
	\label{fig:echogram_raw_pcp}
	\vspace{0.5cm}
\end{figure}

Throughout the selected observation period, narrowband sonar pulses (gated sinusoidal waves with a duration of 1.024 msec) at 38, 120, and 200 kHz were transmitted simultaneously at a 1-sec pinging interval for the first 20 mins of each hour. The volume of the selected echosounder dataset (15.6 GB) is comparable to those from a few days of typical ship-based echosounder survey. The echogram contains distinctive patterns that are easily discerned by eye, facilitating interpretation of the results from matrix decomposition (Fig.~\ref{fig:echogram_raw_pcp}A).

The raw data collected by the echosounder are preprocessed using the Python package \verb|echopype| \cite{Lee2020_echopype} prior to the decomposition analyses. This is a free and open-source software package developed by the authors to allow convenient conversion of manufacturer-specific binary data files into interoperable netCDF or zarr formats. These formats are suitable for storing and computing multi-dimensional data, such as those collected by scientific echosounders. These formats are suitable for storing and computing multi-dimensional data, such as those collected by scientific echosounders. The preprocessing steps include: 1) parsing and saving the \verb|.raw| binary data files to self-described machine-readable netCDF files, 2) performing calibration to obtain volume backscattering strength ($\textrm{S}_\textrm{V}$, units: dB re 1 m$^{-1}$) for each echo sample (each ping) \cite{Simmonds2005}, and 3) obtaining mean volume backscattering strength (MVBS or $\overline{\textrm{S}_\textrm{V}}$) over non-overlapping echogram regions with a depth bin size of 5 m and a temporal bin size of 200 sec. Since this echosounder was not formally calibrated during this deployment, we use manufacturer-supplied parameters stored in the data files to calibrate the echo measurements.

In addition to data from the echosounder, ocean current measurements collected by an acoustic Doppler current profiler (ADCP) from the OOI Oregon Offshore Cabled Benthic Experiment Package (\citeauthor{OOI_CE04OSBP}) are also obtained for joint analysis of the decomposition results.

\subsection{Matrix decomposition}
\subsubsection{Overview}
\label{sect:methods_decomposition_overview}
Matrix decomposition (or factorization) methods seek to represent complex observations through a small set of latent factors in an unsupervised manner while preserving intrinsic structures in the data \cite{Cichocki2009, Bouwmans2016}. These methods leverage the hidden low-dimensional nature of high-dimensional observations and yield latent factors that are usually more tractable and interpretable compared to the original data. These techniques have found wide applications in many scientific and applied fields, including computer vision \cite{Lee1999_nmf, Cichocki2009, Bouwmans2014}, neuroscience \cite{Chen2005, Pnevmatikakis2016}, text mining \cite{Ding2008}, recommender systems \cite{Zhang2006, Koren2009}, deep learning\cite{Oseledets2011}, etc. A notable and illustrative application example is the analysis of human faces using Principal Component Analysis (PCA), which produces a series of ``eigenfaces'' that capture increasingly sophisticated variation of facial features across subjects \cite{Turk1991}. 

Using matrix decomposition methods, here we aim to automatically extract high-level spatio-temporal structures embedded in the echogram of each day and use these patterns as descriptors for temporal changes in the time series. The choice of the daily interval is motivated by the profound influence of daylight cycle on life in the ocean and the significant contribution of diel vertical movements of marine animals in the global biogeochemical cycle \cite{Hays2003, Lehodey2010}. Below we describe the data restructuring procedure and the decomposition formulations used in this study to remove noisy outliers from data, discover distinct echogram patterns, and summarize variations in long-term echosounder time series.

\subsubsection{Restructuring echo data for decomposition}
\label{sect:methods_data_restructuring}
\begin{figure}[!htb]
    \vspace{0.5cm}
	\centering
	\includegraphics[width=0.6\textwidth]{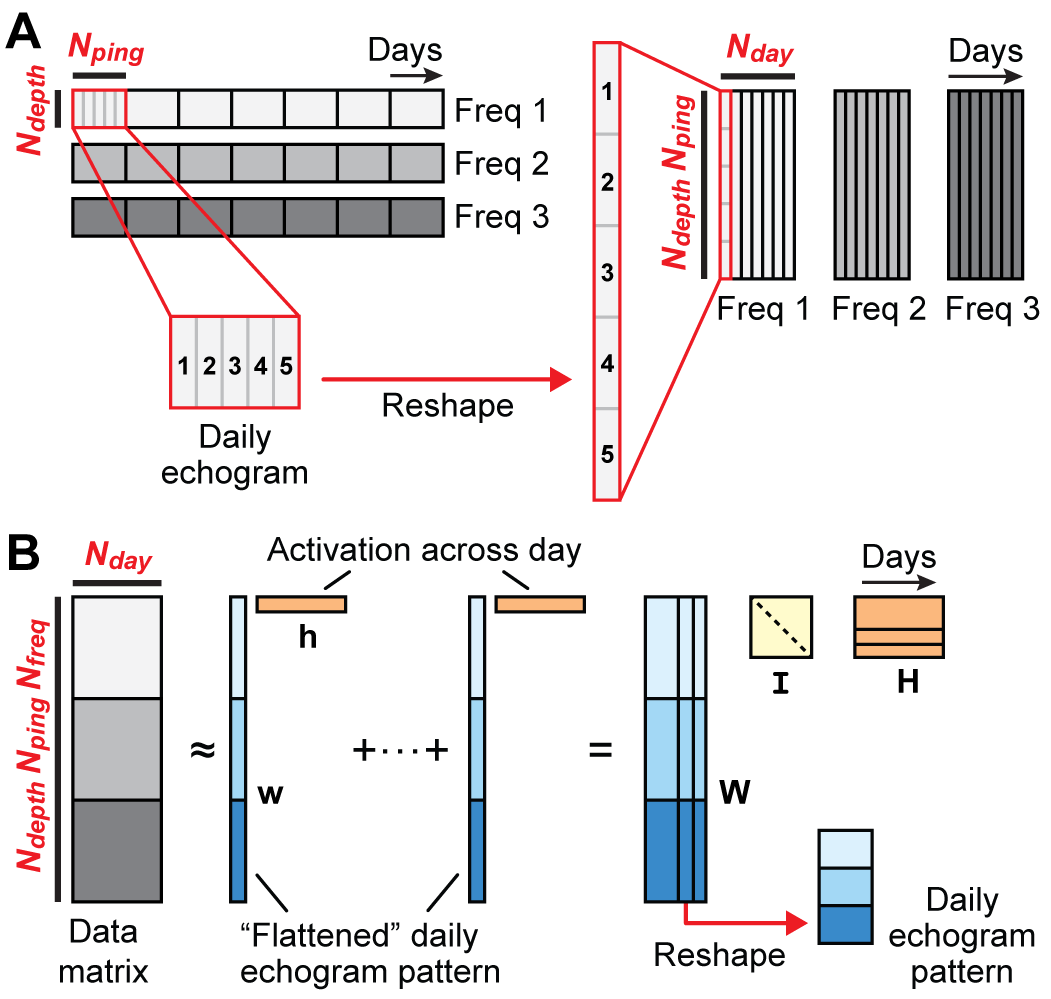}
	\caption{(A) The procedure of restructuring echosounder data for matrix decomposition. Echo observations within individual days are ``flattened'' into long vectors to form a matrix for each frequency. (B) Matrix decomposition, in which the matrices from multiple frequencies are concatenated to form the data matrix. The decomposition results are reshaped back to form daily echogram patterns by reversing the procedure shown in (A). In tsNMF, the formulation includes a temporal smoothness constraint which controls abrupt changes in the activation across day (for each row $\mathbf{h}$ in $\mathbf{H}$). $N_{ping}=144$ is the number of MVBS ping bins within a day, $N_{depth}=37$ is the number of MVBS depth bins, $N_{day}=62$ is the number of days in the selected observation period, and $N_{freq}=3$ is the number of frequencies in the dataset.}
	\label{fig:schematics_decomp}
\end{figure}

The multi-frequency echosounder time series are restructured to allow two-way (matrix) decomposition in this paper. All echo observations from a single frequency of each day are first ``flattened'' and concatenated to form a long vector that contains all echogram pixels within the day. The long vectors from all days in the observation period are then combined horizontally along the day dimension to form a matrix (Fig.~\ref{fig:schematics_decomp}A) for each frequency. The matrices from all frequencies are then stacked vertically to form the final data matrix for decomposition (Fig.~\ref{fig:schematics_decomp}B). Each single-frequency daily echogram is of dimension $N_{depth} \times N_{ping}$, where $N_{depth}$ (=37) is the number of averaged depth bins and $N_{ping}$ (=144) is the number of averaged ping bins of the MVBS data within a day. Therefore, the ``flattened'' long vectors assembled from daily observations are of dimension $N_{depth} N_{ping} \times 1$. The final data matrix is of dimension $N_{depth} N_{ping} N_{freq} \times N_{day}$, where $N_{freq}$ (=3) is the number of frequencies and $N_{day}$ (=62) is the number of days. In addition, all decompositions are performed on the log-transformed MVBS data, due to the well-known challenge of decomposing data with a large dynamic range, such as those commonly encountered in machine listening applications \cite[][see Sec.~\ref{sect:discussion} for more discussion]{Liutkus2015}.

\subsubsection{Outlier removal} 
\label{sect:methods_outlier_removal}
A profiler collocated with the echosounder on the same mooring traversed the water column at irregular time intervals during the observation period and introduced strong interference (indicated by arrows in Fig.~\ref{fig:echogram_raw_pcp}A) to echoes from natural sources. The profiler echo interference needs to be removed from the time series before decomposition analysis to avoid their dominance in the optimization updates. To this end, we employ Principal Component Pursuit (PCP), or robust PCA \cite{Candes2009}, which was designed to separate the data ($\mathbf{X}$) into an exact sum of a low-rank ($\mathbf{L}$) and a sparse ($\mathbf{S}$) component, i.e., $\mathbf{X} = \mathbf{L} + \mathbf{S}$. The development of PCP was driven by the limitations of PCA when dealing with data containing significantly corrupted observations, for which the estimated principal components are easily dominated by the corrupted samples and deviate significantly from the true low-rank structures. The decomposition is obtained by solving the following minimization problem:
\begin{eqnarray}
\label{eq: pcp}
&&\min_{\mathbf{L}, \mathbf{S}} \|\mathbf{L}\|_{*} + \gamma\| \mathbf{S} \|_1,\notag\\
&&\textrm{subject to } \mathbf{L} + \mathbf{S} = \mathbf{X},
\end{eqnarray}
where $\|\cdot\|_{*}$ is the nuclear norm of a matrix, which is the sum of its singular values, and $\|\cdot\|_1$ is the $l_1$ matrix norm, which is equal to the maximum of the $l_1$-norms of its columns ($\| \mathbf{S} \|_1 = \max_j \sum_{i}|s_{ij}|$). 

Compared with the matrix completion problem where the goal is to recover missing entries in the data matrix, PCP handles a similar but broader case in which the locations of the corrupted observations are not known. While one expects some trade-off between the minimization of the nuclear-norm term, which looks for lower dimensional structures, and the $l_1$-norm, which looks for a sparse component, under weak conditions it has been shown by \citet{Candes2009} that such decomposition can be obtained exactly, and, surprisingly, the value of $\gamma$ can be set to $1/\sqrt{\max(dim(\mathbf{X}))}$, making the algorithm parameter-free. While in theory such an exact decomposition may not exist, in practice it has been demonstrated that PCP indeed provides robust decomposition. For example, even though images of faces under different illuminations do not lie in a perfectly low-dimensional space, PCP has been used to recover faces from shadows and occlusions \cite{Candes2009}; similarly, while background in videos is filled with noise and clutter, PCP has been very successful as a background subtraction technique and has found many applications in video surveillance \cite{Bouwmans2014}. It has also been proven that the problem can be solved almost exactly in the presence of small noise \cite{Zhou2010}. The theoretical and practical properties of PCP provide a powerful framework to separate the (sparse) corrupted samples and enable analyses focused on the underlying low-rank structures.

Here we use PCP as a purely data-driven approach to isolate abrupt patches of strong echoes from the profiler and fine echogram granularity (the sparse component) from high-level spatio-temporal structures in the echogram (the low-rank component), which are the focus of subsequent analyses. The decomposition was performed on the restructured MVBS matrix described in Sec.~\ref{sect:methods_data_restructuring}. The PCP decomposition is performed using the Python package \verb|RobustPCA| \cite{ShunChi}.

Beyond the scope of this paper, PCP may be useful in removing electric or acoustic interference commonly observed between unsynchronized active acoustic instruments on ships (e.g., ADCP and echosounders) and in isolating echoes from dense schools of pelagic fish that tend to be spatially and temporally localized on the echogram.

\subsubsection{Additive representation of temporal processes in a long-term echogram} 
\label{sect:methods_tsnmf}
Nonnegative Matrix Factorization (NMF) is a data decomposition alternative to PCA for which both the latent factors and their activation coefficients are constrained to be nonnegative \cite{Paatero1994, Lee1999_nmf}. The nonnegativity constraints make the latent factors (components) purely \textit{additive} building blocks that can be combined to reconstruct the data. Moreover, the components tend to contain primarily localized features, which underlie a more interpretable ``part-based'' reconstruction and representation. The advantage is readily demonstrated by the decomposition of human faces into parts associated with eyes, mouths, etc. shown in \citet{Lee1999_nmf}, as opposed to the eigenfaces from PCA that each contains variations of different parts of the entire face and can sometimes be challenging to interpret.

A common formulation of NMF aims to find low-rank nonnegative matrices $\mathbf{W}$ and $\mathbf{H}$, such that their product minimizes the squared reconstruction error given by the Frobenius norm:
\begin{eqnarray}
\label{eq: nmf}
     &&  \min_{W,H} \left\Vert \mathbf{X} - \mathbf{WH} \right\Vert^2_F,\notag\\
     &&\textrm{subject to } \mathbf{W}_{d,k}\ge0, \mathbf{H}_{k,t}\ge0,
\end{eqnarray}
where $\mathbf{W}$ is of dimension $D\times K$, $\mathbf{H}$ is of dimension $K\times T$, $K\ll D$ is the rank of the decomposition which needs to be predetermined, and $\mathbf{W}_{d,k}$ and $\mathbf{H}_{k,t}$ are the individual entries of the corresponding arrays. We refer to this formulation as the ``traditional NMF'' hereafter. Many variations of this problem exist \cite{Cichocki2009}. For example, additional sparsity and smoothness constraints can be imposed; the Kullback-Leibler divergence, which provides a probabilistic interpretation of the problem \cite{Yang2011}, can be minimized instead of the Frobenius norm of the reconstruction error. 

In this paper we use a regularized temporally smooth Nonnegative Matrix Factorization (tsNMF) formulation to discover patterns ($\mathbf{W}$) embedded in the low-rank MVBS echogram and use the activation of these patterns ($\mathbf{H}$) to describe temporal variations in the echosounder time series. tsNMF modifies the traditional NMF by incorporating in its formulation a temporal smoothness constraint, which controls abrupt changes in the pattern activation sequences and emphasizes the role of sequential information in the decomposition. This is achieved by adding a Tikhonov regularization term of the form $\sum_k\sum_t(\mathbf{H}_{k,t} - \mathbf{H}_{k,t+1})^{2}$, which serves as a discrete equivalent of the squared norm of the derivative of the temporal process. We optimize the following cost function in tsNMF as proposed by \citet{Fabregat2019}:
\begin{equation}
\label{eq:tsnmf}
\begin{split}
    \Psi(\mathbf{W}, \mathbf{H}) 
    &= \left\Vert \mathbf{X} - \mathbf{WH} \right\Vert^2_F 
        + \eta\left\Vert\mathbf{H\mathbf{\Delta}}\right\Vert^2_F \\ 
    & + \lambda \left\Vert \mathbf{W} \right\Vert_1 
        + \beta_{\mathbf{W}}\left\Vert\mathbf{W}\right\Vert^2_F 
        + \beta_{\mathbf{H}}\left\Vert\mathbf{H}\right\Vert^2_F , \\
    & \textrm{subject to } \mathbf{W}_{d,k} \geq 0, \,  \mathbf{H}_{k,t} \geq 0,
\end{split}
\end{equation}
where $\mathbf{\Delta}$ is the difference operator 
\begin{equation}
    \mathbf{\Delta} =
    	\begin{bmatrix}
    		1      & 0      & \dots  & 0 \\
    		-1     & \ddots & \ddots & \vdots \\
    		0      & \ddots & \ddots & 0 \\
    		\vdots & \ddots & \ddots & 1 \\
    		0      & \dots  & 0      & -1
    	\end{bmatrix},
\end{equation}
$\eta$ controls the smoothness of $\mathbf{H}$, $\lambda$ allows tuning for sparsity in $\mathbf{W}$, $\beta_{\mathbf{W}}$ and $\beta_{\mathbf{H}}$ prevent matrices $\mathbf{W}$ and $\mathbf{H}$ from growing too large, and $d$, $k$, $t$ are indices of the matrix elements. We note that when all parameters $\eta$, $\lambda$, $\beta_\mathbf{W}$ and $\beta_\mathbf{H}$ are set to zero, we obtain the traditional NMF which simply minimizes the reconstruction cost subject to the nonnegativity constraints. To understand the importance of the temporal constraint, note that in the traditional NMF a permutation of $\mathbf{X}$ along the dimension $T$ would not change the result.

In our context of sonar data analysis, $\mathbf{X}$ is the low-rank MVBS echogram $\mathbf{L}$ from the PCP decomposition (with dimensions $D = N_{depth} N_{ping} N_{freq}$ and $T = N_{day}$), $\mathbf{W}$ contains the latent factors (patterns), $\mathbf{H}$ contains the patterns' activation coefficients, and the rank $K$ (which equals the number of patterns) is selected based on the procedure described in Sec.~\ref{sect:methods_param_selection}. Specifically, column $\mathbf{w}_k$ in $\mathbf{W}$ encodes a particular daily echogram pattern, with the corresponding row $\mathbf{h}_k$ in $\mathbf{H}$ capturing the activation of this pattern in the observation period (Fig.~\ref{fig:schematics_decomp}B). The summation of the outer products of the $K$ pairs of $\mathbf{w}_k$ and $\mathbf{h}_k$ reconstructs the low-rank MVBS data. 

Since values in the MVBS data matrix are negative in absolute physical units, we perform nonnegative decomposition by first subtracting the minimum of the entire matrix and adding the minimal value back for reconstruction. This approach is justified by our primary focus on extracting high-level patterns in the data, rather than physics-based inversion of scatterer numerical density (see Sec.~\ref{sect:discussion} for more discussion).

The optimization is solved by a Proximal Alternating Linearized Minimization (PALM) algorithm, which was proposed by \citet{Bolte2014} as an algorithm with favorable convergent properties for a wide class of constrained matrix decomposition problems and later derived by \citet{Fabregat2019} for the tsNMF formulation presented here. We implemented this method in the Python package \verb|time-series-nmf| \cite{Staneva2020}.

\subsubsection{Selection of decomposition parameters}
\label{sect:methods_param_selection}
In this work we select the rank, regularization parameters, and initialization of $\mathbf{H}$ and $\mathbf{W}$ in tsNMF based on empirical analysis of the decomposition outcome. Finding the intrinsic low dimension of a data set has long been a challenging task for many unsupervised techniques, including clustering \citep{Mirkin2011}. We use the variation of the mean squared error (MSE) of the reconstruction \cite{Hutchins2008, Frigyesi2008} and the cophenetic correlation coefficient \cite{Brunet2004} across increasing rank as guides for rank selection. The former provides an indication of the rank beyond which no significant non-random structures in the data are captured by the decomposition, and the latter provides a measure for decomposition stability across multiple randomly initialized runs. Due to the requirement to assign a given observation day to only one daily echogram pattern in the calculation of the cophenetic correlation coefficient, which conflicts with the fact that it is the joint activation of multiple patterns which reconstructs the data, here we base our rank selection primarily on MSE variation and use the variation of cophenetic correlation coefficient only as a reference. The smoothness parameter $\eta$ is selected based on the L-curve concept \cite{Oraintara2000, Mirzal2014}, which seeks to balance the reconstruction error and the smoothness regularization cost in the decomposition. Details of these metrics are described in the \hyperref[sect:appendix]{Appendix}. 

We found that the exact values of $\lambda$, $\beta_{\mathbf{H}}$ and $\beta_{\mathbf{W}}$ do not impact the decomposition results dramatically, as long as the values are chosen within a reasonable range such that the respective terms do not entirely dominate the cost and diminish the contributions of the reconstruction and the smoothness regularization terms. Specifically, increasing $\lambda$ does make the resulting components $\mathbf{W}$ sparser (which in turn causes the activations $\mathbf{H}$ to be smoother), but does not affect the primary structures in $\mathbf{H}$ and $\mathbf{W}$. Therefore, we set $\lambda=0$ when presenting our results for better interpretability. \citet{Fabregat2019} suggested setting $\beta_{\mathbf{H}}$ and $\beta_{\mathbf{W}}$ to arbitrary positive numbers and used $0.1$ in their demonstration. However, we opted for using $\beta_{\mathbf{H}} = \beta_{\mathbf{W}} = 0$ as we found no practical differences for $0 \leq \beta_{\mathbf{H}}, \beta_{\mathbf{W}} \lesssim 0.1$. 

The tsNMF optimization iteration stops when the rate of cost decrease is smaller than 0.5\% of the running average of the rate of cost decease over a window of 5 iterations. The components $\mathbf{W}$ and activations $\mathbf{H}$ shown in all figures are obtained by running 320 randomly initialized decompositions and selecting the run with the lowest representative cost.

\subsection{Summarizing variations in a long-term echosounder time series} \label{sect:methods_similarity}
The activation of daily echogram patterns from matrix decomposition provides a compact representation that is useful for summarizing temporal processes in a long-term echosounder time series. To assess changes in the overall echogram structure, we compute the Euclidean distance between the pattern activation for pairs of days and perform hierarchical agglomerative clustering using Ward's minimum variance method, in which clusters are merged based on the criterion of minimizing the sum of squared differences within all clusters \citep{Ward1963}. The distance matrix and clustering outcome allow convenient identification of the transitions of overall echogram structure throughout the observation period.

\section{Results}

\subsection{Outlier-free low-rank echogram} 
\label{sect:results_outlier_removal}
PCP successfully decomposes the MVBS echogram into a low-rank component that contains the high-level spatio-temporal features in the data (Fig.~\ref{fig:echogram_raw_pcp}B) and a sparse component consisting of abrupt echo patches and fine-grained details (Fig.~\ref{fig:si_pcp_sparse}). The low-rank component is significantly ``cleaner'' compared to the original MVBS echogram while still retains the high-level structures we aim to capture and describe. The dominant features in the sparse component match the echo traces of the profiler, which corrupt the original MVBS time series. This is achieved without prior knowledge of the profiler movement schedule, which was in fact irregular during the observation period due to maintenance needs. In addition, faint DVM patterns are observed in the background of the sparse component. This is likely due to the fact that there are daily variations in the high-level patterns (which are themselves sparse) as well as noise, i.e. the MVBS data are not a perfectly identifiable sum of $\mathbf{L}$ and $\mathbf{S}$. Importantly, recall that there are no parameters to tune in the PCP algorithm (Sec.~\ref{sect:methods_outlier_removal}) and thus we have traded some flexibility in the decomposition for the stability of the implementation, which is needed for long-term studies where little is known about the properties of individual processes and the distributions of noise and outliers. If PCP is to be used as a tool for removing noise from calibrated echosounder data with an objective of biomass estimation, further investigation into its algorithmic implementation is needed to ensure that the sparse daily variabilities in the data can be accounted for in the downstream processing.

\subsection{Daily echogram patterns as building blocks of long time series} 
\label{sect:results_nmf}
\begin{figure}[!htb]
	\centering
	\includegraphics[width=1\textwidth]{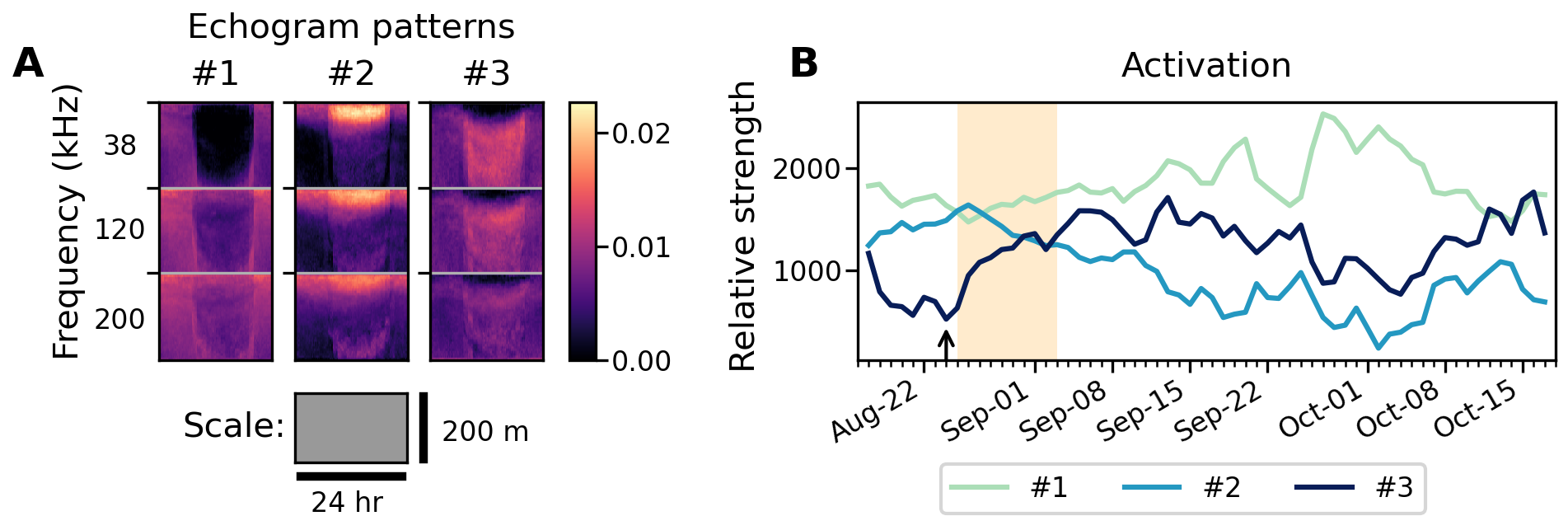}
	\caption{Decomposition results of tsNMF at rank=3. (A) Daily echogram patterns. The patterns across all three frequencies are displayed together because the decomposition is performed on concatenated multi-frequency feature vectors (Fig.~\ref{fig:schematics_decomp}, Sec.~\ref{sect:methods_data_restructuring}). The gray square at the bottom shows the spatial and temporal scale of the patterns. (B) Corresponding activation sequences of the daily echogram patterns. The arrow indicates the period of low activation of component \#3 also indicated in Fig.~\ref{fig:echogram_raw_pcp}B. The yellow shading indicates the time period included in the dashed box in Fig.~\ref{fig:echogram_raw_pcp}B. Note that the norm of each daily echogram pattern $\left\Vert\mathbf{w}\right\Vert_2$ is factored out and multiplied with the corresponding activation sequence to facilitate interpretation.}
	\label{fig:results_decomp}
	\vspace{0.5cm}
\end{figure}

The low-rank MVBS echogram from PCP is decomposed by tsNMF (rank=3, $\eta=500000$) into three distinct daily echogram patterns with corresponding activation sequences throughout the observation period (Fig.~\ref{fig:results_decomp}). Three is the lowest rank at which the decomposition produces clearly distinguishable daily echogram patterns and at the same time captures the majority of the structures in the data (for additional details see the \hyperref[sect:appendix]{Appendix} and Fig.~\ref{fig:si_rank_4}).

The three daily echogram patterns collectively represent a repertoire of daily echo energy distributions that jointly reconstruct the low-rank MVBS time series, demonstrating the ``parts-based representation'' property of nonnegative decomposition (see Sec.~\ref{sect:methods_tsnmf}). Pattern \#1 resembles the diel vertical migration (DVM) behavior widely observed in the global ocean \cite{Hays2003}. The trend of increasing echo strength with increasing frequency in this pattern is consistent with the echo spectral feature of zooplankton-like scatterers at the observation frequencies. The activation sequence of Pattern \#1 indicates relatively high contribution throughout the observation period, with enhanced activity between early September to early October, which corresponds to visibly intensified DVM during the same time period in the MVBS echogram. 

Pattern \#2 consists of an intense sub-surface layer and a fainter DVM-like sub-pattern at a shallower depth than that in Pattern \#1. The two sub-patterns exhibit different frequency-dependent trends on echo strength: the echo strength variation of the sub-surface layer has a weak positive slope during the night time hours and a weak negative slope during the day time hours (toward the outer and middle vertical section of the pattern, respectively), whereas the DVM-like sub-pattern exhibits a clear positive slope with increasing frequency. The activation of Pattern \#2 captures the strong sub-surface layer and its gradual disappearance in the first one-third of the MVBS echogram. However, the gradual sinking and dissipation of the sub-surface layer cannot be completely captured by the current decomposition model, which assumes fixed daily echogram patterns that do not change much over time (see Sec.~\ref{sect:discussion} for more discussion). We further observe that some echogram features are best represented jointly by two patterns. For example, the gradual depth variation of the DVM in late August (highlighted by the dashed box in Fig.~\ref{fig:echogram_raw_pcp}B) is captured by the combination of Pattern \#1 and the DVM-like sub-pattern in Pattern \#2, and the positive slope of echo strength variation across frequency for the sub-surface layer requires both Pattern \#1 and \#2 to reconstruct.

Pattern \#3 represents an aggregation of scatterers that appear in the mid to upper water column during daytime, with a trend of decreasing echo strength over increasing frequency that is consistent with fish-like scatterers. While its activation is more difficult to discern than the other two patterns, its inactivation produces noticeable changes on the echogram: the mid to upper water column appears significantly ``emptier,'' especially at 38 kHz, when Pattern \#3 activation is low around August 24 and from late September to early October.

\begin{figure}[!htb]
	\centering
	\includegraphics[width=0.55\textwidth]{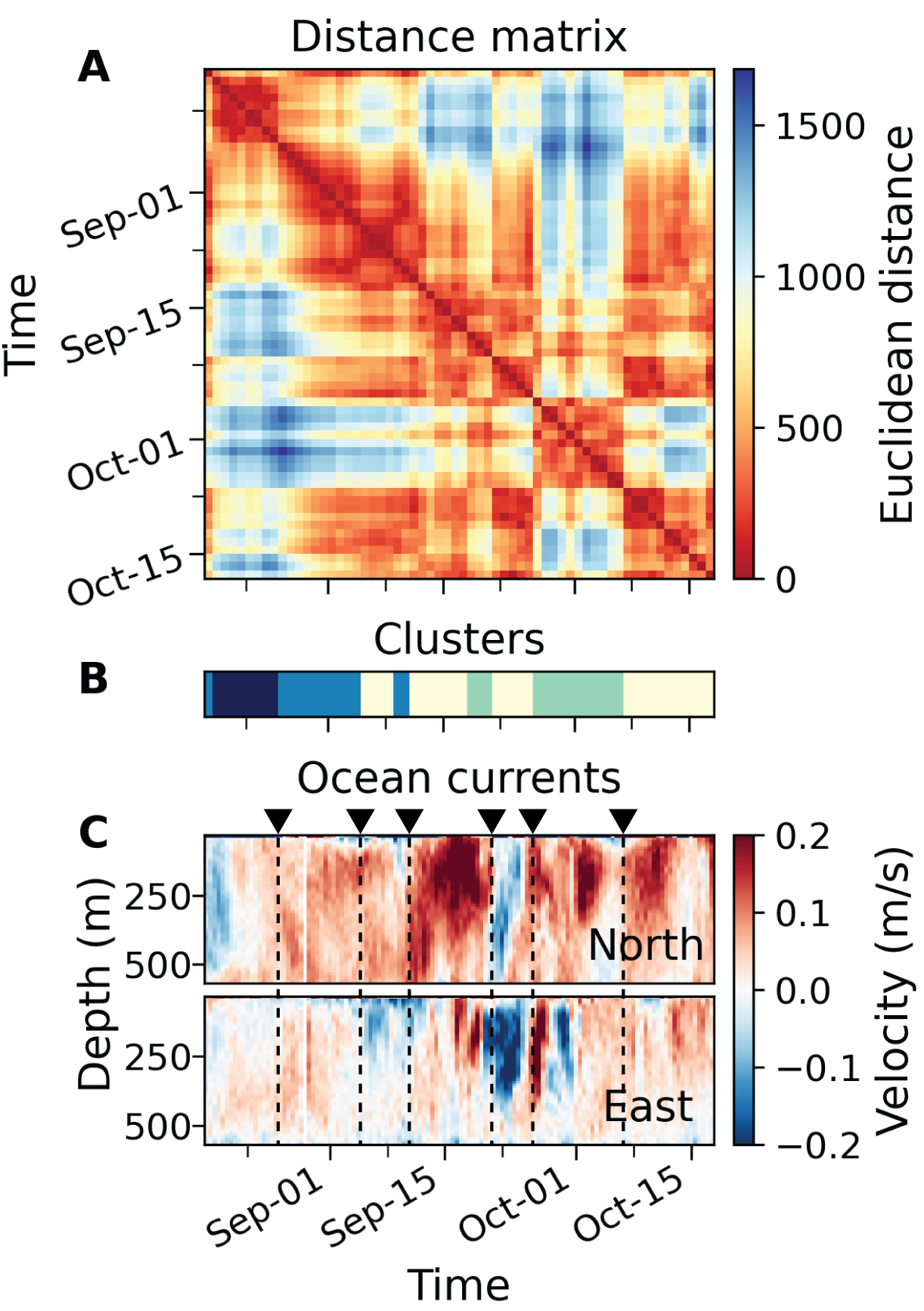}
	\caption{(A) Distance matrix computed by pairwise Euclidean distance of the pattern activation between days. (B) Transition between four clusters of pattern co-activation derived from the distance matrix. Cluster features are shown in Fig.~\ref{fig:si_clustering_dendrogram}. (C) Ocean currents in the North-South (top) and East-West (bottom) directions measured by an ADCP geographically collocated with the echosounder. The arrows and dash lines are drawn according to the cluster transition time points at which the transition coincides with changes in the direction or magnitude of ocean currents. The horizontal axes of all panels are identical.}
	\label{fig:distance_matrix}
\end{figure}

\subsection{Temporal changes of echogram structure and the environment} \label{sect:results_similarity}
The distance matrix calculated based on the activation strengths of daily echogram patterns provides a summarization of the multiple transitions of overall echogram structure during the 62-day observation period (Fig.~\ref{fig:distance_matrix}A). Along the matrix diagonal, groups of consecutive days that share similar co-activation of echogram patterns are clearly identifiable as reddish ``blocks.'' The off-diagonal entries show similarity between temporally more distant observation days and reveal recurring co-activation of the daily echogram patterns. The echogram structural changes are also illustrated by the multiple transitions between four clusters of pattern co-activation identified by agglomerative clustering based on pairwise distance between days (Fig.~\ref{fig:distance_matrix}B and Fig.~\ref{fig:si_clustering_dendrogram}). Pairing the distance matrix and the cluster transition diagram with ADCP measurements further reveals interesting coincidence at multiple time points between overall echogram structural changes and prominent direction reversal or magnitude variation of ocean currents at the upper 300 m of the water column (indicated by the arrows and dash lines in Fig.~\ref{fig:distance_matrix}C). For example, the pattern activation cluster transitions at September 3, 10, 22, and 24 coincide with direction reversal in one or both of the eastward and northward components of the currents. The cluster transitions at August 25 and October 7 additionally coincide with an increase of current velocity in the lower and upper water column, respectively.

\section{Discussion} 
\label{sect:discussion}

In this paper we present automatic extraction and compact representation of temporal processes in a long-term echosounder time series via matrix decomposition. We show that unsupervised decomposition techniques are capable of transforming complex echo observation into low-dimensional components that are more tractable and interpretable than the original data. The decomposition methods automatically remove noisy outliers and capture high-level spatio-temporal echogram structures through time-varying linear combination of a small number of ecologically relevant daily echogram patterns. The implication of our results is profound: this approach is entirely data-driven and does not assume knowledge of the type of scatterers or sources of interference present in the data. Such an approach is fundamentally different from previous echo analysis methods that rely heavily on handcrafted rules built from past experience, limited biological ground truth information, and observer assumptions. The methodology we present here is therefore more flexible and adaptable when applied to data collected from geographical regions or environmental conditions for which scarce prior information exists. This type of data is now prevalent due to the broad availability of echosounder-equipped autonomous ocean observing platforms and as a result of the rapidly changing climate.

Our results show that the parameter-free, automatic PCP decomposition significantly reduces the dimensionality of the MVBS echogram (Fig.~\ref{fig:echogram_raw_pcp}) and enables the subsequent tsNMF decomposition analysis. By restructuring data according to the daily interval (Fig.~\ref{fig:schematics_decomp}), we exploit the intrinsic regularity in the data through PCP to remove the echo interference from the water column profiler. Without this crucial step, the profiler echo traces would have dominated the tsNMF optimization updates due to their significant echo strengths. In addition, while much fine echogram granularity is assigned to the sparse component by PCP, the low-rank MVBS echogram retains the high-level spatio-temporal structures intrinsic to the original data that are important for joint analysis of echosounder observation with other ocean environmental parameters.

The results also show that a temporally regularized nonnegative decomposition through tsNMF provides a tractable and interpretable dissection of the echosounder time series (Fig.~\ref{fig:results_decomp}). The decomposition discovers daily echogram patterns that each represents a distribution of echo energy across time (hour of day) and space (depth), with varying daily activation throughout the observation period. For biological scatterers, such grouping is based on \textit{behavior} (vertical movements within a day) and does not necessarily correspond to specific taxonomic group of organisms that have been the focus of many echo classification methods (see Sec.~\ref{sect:introduction}). This behavioral binding can be ecologically significant, since spatial and temporal co-occurrence of organisms within daily movements suggests trophic functional grouping that mediates energy transfer throughout the water column \cite{Lehodey2010}.


\subsection*{Modeling frequency domain information}
Interestingly, while frequency domain information is not explicitly modeled in tsNMF, the constraint on temporal smoothness of pattern activation encourages biological (and therefore spectral) coherence within each pattern, since movements of animal aggregations are continuous in time and space. Recall that spectral characteristics of echoes vary significantly depending on the size, shape, orientation, and material properties of the scatterer. In a preliminary study \cite{Lee2019_tensor} we examined the effects of explicitly modeling frequency domain information via tensor decomposition, which extends matrix decomposition to a multi-way component analysis. While the nonnegative tensor decomposition yielded three daily echogram patterns that resemble the general distribution of echo energy in Pattern \#1 and \#2 discovered by tsNMF here, the spectral dependency within each pattern is different. Further investigation revealed that the reconstruction error is substantially higher for tensor decomposition, rendering its results less reliable and less interpretable. This is likely due to the combination of 1) the restrictive form of Kruskal tensor decomposition we used, which requires coherent variation across frequency for \textit{all pixels} within a single pattern but at the same time does not allow interaction across patterns in the reconstruction \cite{Cichocki2009}, and 2) the small number of frequency features ($N_f=3$) in the dataset, which limits the total number of patterns (rank) allowed in the decomposition. Future investigations using data with richer spectral features, such as those collected by broadband echosounders, would be more appropriate for evaluating the utility of tensor decomposition in modeling frequency domain information in the decomposition.

\subsection*{Relation to environmental data}
The pattern activation sequences from tsNMF and the derived distance matrix summarize both similarities and changes in the echosounder time series, providing an explicit avenue for joint analysis of acoustic observation and ocean environmental data. By aligning ocean currents measured by ADCP with the distance matrix and activation clusters, a clear pattern emerges that associates changes in the echogram structure to changes of the direction or magnitude of ocean currents at multiple time points throughout the observation period (indicated by arrows in Fig.~\ref{fig:distance_matrix}C). These associations are ecologically plausible, since mid-trophic animals can be advected by strong currents, and their vertical movement behaviours may also be modulated by changes in the ambient environment. While detailed interpretation of these observations is beyond the scope of this paper, we note that strong currents (up to $\sim$0.2 m/s) could be biologically significant in this region \cite{Keister2009, Wu2014}. In addition, even though conventional echogram summary statistics may show association with a subset of the observed current changes (e.g., Fig.~\ref{fig:si_summary_stat}), the statistics do not provide easily interpretable summarization of the actual echogram structure in the same way the tsNMF patterns and activations do. Unlike the fixed, handcrafted definitions of the summary statistics, our \textit{data-driven} methodology further ensures that the automatically extracted descriptors (the patterns) change adaptively with intrinsic structures in the data. Using an example dataset consisting of 62 days of observation, we illustrate the synoptic summarizing power of unsupervised decomposition that will likely play a crucial role in extracting information from much longer time series that would be difficult to analyze manually.

\subsection*{Comparison with PCA}
Our results demonstrate the advantage of interpretability from imposing nonnegative constraints in decomposition analysis of echosounder data. In comparison with PCA, the purely additive linear combination of low-rank patterns makes it straightforward to attribute changes in the long-term echogram to temporally varying contributions from different daily echogram patterns. For example, while the appearance of the first two PCA components resembles the first two tsNMF patterns, their contributions to the observed MVBS echogram are much more difficult to interpret due to the intermingled positive and negative entries in both the patterns and the activation sequences (Fig.~\ref{fig:si_pca}). PCA could however useful for analyzing anomaly with respect to a mean pattern, as has been shown in \citet{Parra2019}.

\subsection*{Rank and parameter selection}
The selection of hyperparameters in the decomposition, such as the rank and the smoothness parameter, is a topic that requires attention in future application of similar methods. In this work we use intrinsic properties of the data to select these hyperparameters: we select the minimum rank that yields distinctive daily echogram patterns with moderate reconstruction error, and choose a smoothness parameter that resides at the corner of the L-curve, which balances the relative contribution of reconstruction error and smoothness regularization cost in the decomposition optimization (Sec.~\ref{sect:methods_param_selection}). However, the choice of these parameters may also depend on the exact context of analysis. For example, the choice of rank, or the number of echogram patterns, may be driven by strong prior expectation on the number of organism types in the observed area, if such information is available. When echosounder data are analyzed in conjunction with other oceanographic measurements over a large time period, the choice of the rank and the smoothness parameter may also be influenced by how well the decomposition captures echogram structures that temporally co-vary with other ocean variables.

\subsection*{Future development and implications}
Our results further help pinpoint specific areas of improvement for matrix decomposition analysis of long-term echosounder time series. First, while tsNMF amends the traditional NMF model by imposing a constraint on temporal smoothness, incorporating spatial smoothness regularization (over depth and time within a day) would likely improve the biological coherence within each pattern. Second, many high-level echogram patterns are transient in time and therefore would be more appropriately captured by a temporally adaptive formulation that allows gradual intrinsic changes \textit{within} the daily echogram patterns (in addition to the time-varying activation strengths \textit{across} day). For example, while Pattern \#2 captures the strong sub-surface layer in the data, the gradual sinking and dissipation of this layer from August 29 to September 10 is only partially captured as a decrease of activation of this fixed pattern. Similar gradual changes are expected for the timing and depth of DVM across seasons for longer time series. In fact, allowing temporally adaptive factors is a challenging problem across diverse domains that many dynamic decomposition formulations aim to address \cite{Saha2012, Aravkin2016_social, Mohammadiha2015, Chen2015}. Third, data from ocean observing platforms are frequently subject to gaps in the time series due to instrumentation or communication problems. While short data gaps can likely be handled by excluding the missing entries in tsNMF and interpolating the resulting activation sequences, a full methodological development to model the temporal evolution of pattern activations will be needed to adequately address this issue. In addition, the reconstruction error in tsNMF is evaluated through the Frobenius norm, which corresponds to an assumed Gaussian noise model. A different model is likely needed to tackle the challenge of linear decomposition of data with a large dynamic ranges \cite{Liutkus2015}. This has been an outstanding problem in music and audio analysis, for which data with large dynamic range are similarly common. In this study we perform PCP and tsNMF analyses on log-transformed MVBS data based on the common practice in echo processing \citep[e.g.,][]{DeRobertis2010_multifreq}. However, a linear decomposition of echo data in the linear domain could open the door for direct physics-based inversion of the decomposition results for important biological quantities, such as biomass, provided that accurate echosounder calibration information is available. 

In this paper we show that unsupervised matrix decomposition methods are promising techniques that can automatic extract ecologically relevant information and provide synoptic insights from long echosounder time series given minimal prior information. This work forms the basis for future development of effective methodologies that integrate echosounder observation into large-scale, autonomous ocean observation strategies.

\section*{Data availability and code repository}
All data used in this work are openly available on the OOI Raw Data Archive: \url{https://rawdata.oceanobservatories.org/files/}. The ADCP data can additionally be accessed programmatically through the OOI machine to machine interface:  \url{https://oceanobservatories.org/ooi-m2m-interface/}. All analyses performed in this study can be accessed in the GitHub repository: \url{https://github.com/leewujung/ooi-echo-matrix-decomposition}.


\section*{Acknowledgement}
We thank Aleksandr Aravkin and Julie Keister at the University of Washington and Dezhang Chu at the NOAA Northwest Fisheries Science Center for their helpful suggestions on this work. WJL and VS were supported by NSF award \#1849930. VS was also supported by the Gordon and Betty Moore and Alfred P. Sloan foundations Data Science Environments (MSDSE).


\section*{Appendix}
\label{sect:appendix}

\subsection{Variation of reconstruction error with increasing rank}
\begin{figure}[!htb]
	\centering
	\includegraphics[width=0.5\textwidth]{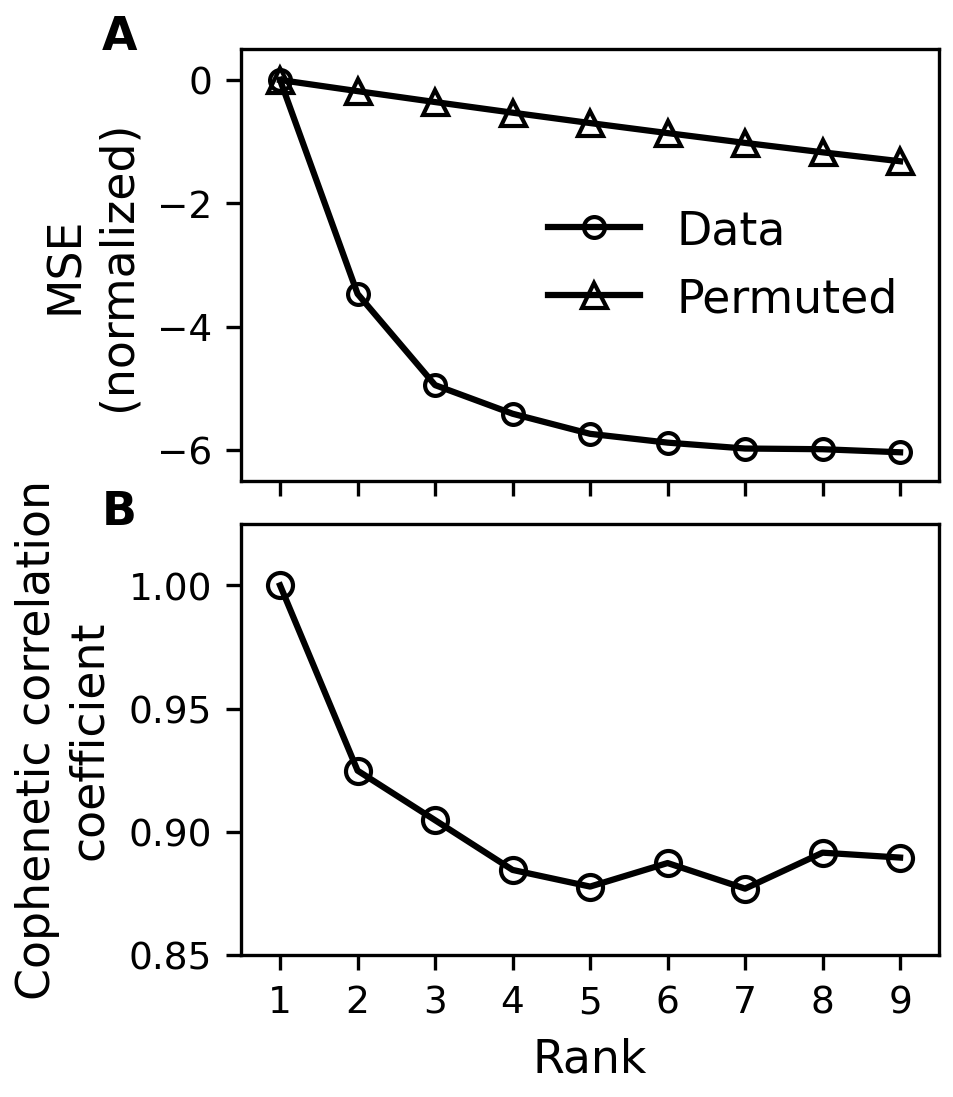}
	\caption{Rank selection. (A) Variation of the MSE of reconstruction with increasing rank for tsNMF decomposition of the low-rank MVBS data and its permutation across the observation days (across columns in Fig.~\ref{fig:schematics_decomp}). This comparison provides a guide for selecting the rank beyond which no significant data structures are captured by the decomposition \cite{Hutchins2008, Frigyesi2008}. Each symbol is the median MSE of all 320 randomly initialized runs. Each curve is normalized by subtracting the median MSE at rank=1. (B) Variation of the cophenetic correlation coefficient with increasing rank. This coefficient provides a measure for decomposition stability \cite{Brunet2004}. Due to the requirement to assign a given observation day to only one daily echogram pattern in the calculation of cophenetic correlation coefficient, which conflicts with the fact that the data is reconstructed by the joint activation of multiple patterns, here we base our rank selection primarily on MSE variation and use the variation of cophenetic correlation coefficient only as a reference. See Sec.~\ref{sect:methods_param_selection} and Sec.~\ref{sect:results_nmf} for detail.}
	\label{fig:si_rank_sel}
	\vspace{0.5cm}
\end{figure}

The variation of the mean squared error (MSE) of NMF reconstruction was used to estimate the number of components in a number of previous studies \cite[e.g.,][]{Kim2003, Frigyesi2008, Hutchins2008}. The method involves comparing the decrease of MSE across increasing rank between the data and a randomized permutation of the data (Fig.~\ref{fig:si_rank_sel}A), and selects the rank at which the MSE decrease starts to level or when the slope of MSE reduction is smaller than that of the permuted data. The rationale is that when the number of components (the specified rank) exceeds the true rank of the data, the extra components would not capture meaningful structures in the data, and hence the MSE reduction would either become less significant or approach the approximately linear slope of a randomly permuted data set, in which the original structure in the data has been disrupted. In this study we permute the low-rank MVBS data matrix across the $N_{day}$ observations (across columns in Fig.~\ref{fig:schematics_decomp}), so that the variance within each pixel (a specific hour-depth combination) of a daily echogram is preserved but its associations with other pixels are disrupted.

\subsection{Cophenetic correlation}
Cophenetic correlation was originally introduced in the context of hierarchical clustering for quantitative comparison of dendrograms \cite{Sokal1962} by measuring how truthfully a dendrogram preserves the pairwise distance between individual observations (samples), and has since become a popular method for determining the number of clusters in clustering problems. It was later used as a method to determine the number of components in NMF based on the stability of multiple randomly initiated NMF runs \cite{Brunet2004}. Specifically, the decomposition can be interpreted as a clustering operation, with each component representing a cluster and each observation assigned to the cluster for which the corresponding activation is highest. While the exact order of components in NMF may not be consistent over multiple runs, the more stable the decomposition, the more likely a pair of similar observations would be assigned to the same cluster. The cophenetic correlation coefficient is a measure of the dispersion of the consensus of the cluster assignment across runs and can be calculated for a set of ranks to select the number of clusters (the rank) (Fig.~\ref{fig:si_rank_sel}B). \citet{Brunet2004} suggested to select the rank at which the cophenetic correlation coefficient begins to fall.

The cophenetic correlation coefficient is computed in the following way: For each NMF run, a connectivity matrix $\mathbf{C}$ is an indicator matrix, within which each entry indicates whether a pair of observations belong to the same cluster: 
\begin{equation}
\mathbf{C}_{ij} = 
\begin{cases} 1 \textrm{, if the {${i}$th} and {${j}$th} observations belong to the same cluster,}\\
0 \textrm{, otherwise}.
\end{cases}
\end{equation}
The consensus matrix $\bar{\mathbf{C}}_{ij}$ for an ensemble of NMF runs is the fraction of the times a pair of observations belong to the same cluster and is equal to the average of the connectivity matrix over all runs. Following the suggestion by \citet{Frigyesi2008}, we use a weighted version of the consensus matrix
\begin{equation}
\bar{\mathbf{C}}_{ij} = \sum_{n=1}^N \mathbf{C}_{ij}(n) \frac{w_n}{\sum_{n=1}^N w_n},
\end{equation}
where $n$ is the run number and $w_n = ((max(e)-e(n))/(max(e)-min(e))$, in which $e(n)$ is the MSE of the $n$th run. This weighting emphasizes contributions from runs with smaller MSE. The closer $\bar{\mathbf{C}}_{ij}$ is to $1$, the higher the tendency the $i$th and the $j$th observations are clustered together. The consensus matrix can then be converted to a dissimilarity matrix $\mathbf{D}_{ij} = 1 - \bar{\mathbf{C}}_{ij}$. The cophenetic distance between two observations is equal to the height of the link between them from a hierarchical clustering based on the dissimilarity matrix. The cophenetic correlation coefficient is obtained by computing the Pearson correlation between the original dissimilarity of the observations and the corresponding cophentic distance. The cophenetic correlation coefficient is $1$ when the cluster assignments are identical for all runs.

\subsection{The L-curve}
\begin{figure}[!htb]
	\centering
	\includegraphics[width=0.5\textwidth]{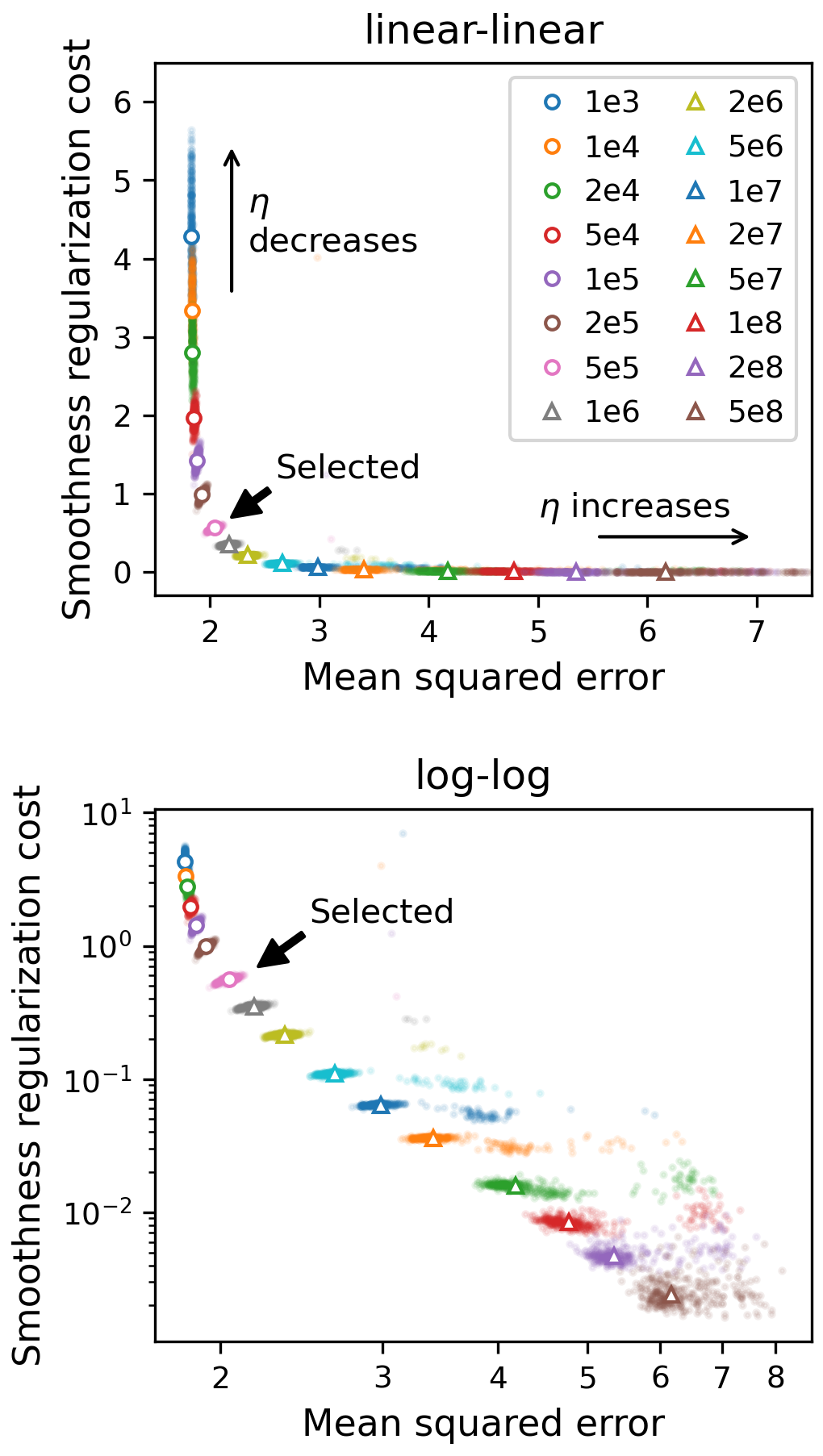}
	\caption{The co-variation of the mean squared reconstruction error and the smoothness regularization cost resulting from changes of the smoothness parameter $\eta$ (the ``L-curve'') on a linear-linear and a log-log plot. Each colored cloud of points represents 320 tsNMF runs under a specific $\eta$ value, with the rank $K=3$ and $\lambda=\beta_\mathbf{H}=\beta_\mathbf{W}=0$. The symbols are median values of all runs. In this paper we follow the recommendation from the literature to select $\eta = 500000$ that corresponds to the ``corner'' of the L-curve.}
	\label{fig:si_L_curve}
	\vspace{0.5cm}
\end{figure}

The tsNMF formulation used in this work is a form of regularized decomposition for which the regularization parameter can be tuned to fit the needs of the application. Specifically, the value of the smoothness parameter $\eta$ controls the balance between the reconstruction error and the smoothness regularization cost in the decomposition (the first and the second term in eq.~(\ref{eq:tsnmf}), respectively). As $\eta$ increases, the decomposition transitions from a regime dominated by the reconstruction error to a regime dominated by the smoothness regularization cost, forming an ``L-curve'' that has been widely recognized in the literature \cite[e.g.,][]{Oraintara2000, Mirzal2014}. In this paper, we follow the recommendation from these previous studies and select $\eta = 500000$, which corresponds to the ``corner'' of the L-curve resulting from rank=3 tsNMF runs over a broad range of parameter values (Fig.~\ref{fig:si_L_curve}). The effects of adjusting $\eta$ on the decomposition results are shown in Fig.~\ref{fig:si_sm_effects}.

\newpage
\section*{References}

\bibliography{ooi-nmf_202009}

\begin{thebibliography}{67}
\def\enquote#1{``#1,''}
\def\plainquote#1{``#1''}
\expandafter\ifx\csname natexlab\endcsname\relax\def\natexlab#1{#1}\fi
\providecommand{\dourl}[1]{\href{http://#1}{\nolinkurl{#1}}}
\providecommand{\bibinfo}[2]{#2}
\providecommand{\noopsort}[1]{}
\providecommand{\switchargs}[2]{#2#1}
  \def\eatspace #1{#1}

\bibitem[{Aravkin \emph{et~al.}(2016)Aravkin, Varshney, and
  Yang}]{Aravkin2016_social}
\bibinfo{author}{Aravkin, A.~Y.}, \bibinfo{author}{Varshney, K.~R.},  and
  \bibinfo{author}{Yang, L.} (\textbf{\bibinfo{year}{2016}}).
  \enquote{\bibinfo{title}{{Dynamic matrix factorization with social
  influence}}} in \emph{\bibinfo{booktitle}{IEEE International Workshop on
  Machine Learning for Signal Processing, MLSP}}, \bibinfo{address}{Salerno,
  Italy}, Vol. 2016-Novem, \dourl{http://arxiv.org/abs/1604.06194},
  \dodoi{10.1109/MLSP.2016.7738846}.

\bibitem[{Benoit-Bird and Lawson(2016)}]{Benoit-Bird2016}
\bibinfo{author}{Benoit-Bird, K.~J.},  and \bibinfo{author}{Lawson, G.~L.}
  (\textbf{\bibinfo{year}{2016}}). \enquote{\bibinfo{title}{{Ecological
  insights from pelagic habitats acquired using active acoustic techniques}}}
  \bibinfo{journal}{Annual Review of Marine Science} \textbf{8}(1),
  \bibinfo{pages}{463--490}, \dodoi{10.1146/annurev-marine-122414-034001}.

\bibitem[{Bolte \emph{et~al.}(2014)Bolte, Sabach, and Teboulle}]{Bolte2014}
\bibinfo{author}{Bolte, J.}, \bibinfo{author}{Sabach, S.},  and
  \bibinfo{author}{Teboulle, M.} (\textbf{\bibinfo{year}{2014}}).
  \enquote{\bibinfo{title}{{Proximal alternating linearized minimization for
  nonconvex and nonsmooth problems}}} \bibinfo{journal}{Mathematical
  Programming} \textbf{146}(1-2), \bibinfo{pages}{459--494},
  \dodoi{10.1007/s10107-013-0701-9}.

\bibitem[{Bouwmans \emph{et~al.}(2016)Bouwmans, Aybat, and
  Zahzah}]{Bouwmans2016}
\bibinfo{editor}{Bouwmans, T.}, \bibinfo{editor}{Aybat, N.~S.},  and
  \bibinfo{editor}{Zahzah, E.-h.}, eds. (\textbf{\bibinfo{year}{2016}}).
  \emph{\bibinfo{title}{{Handbook of Robust Low-Rank and Sparse Matrix
  Decomposition}}} (\bibinfo{publisher}{Chapman and Hall/RC Press}),
  \dourl{https://www.taylorfrancis.com/books/9781498724630}.

\bibitem[{Bouwmans and Zahzah(2014)}]{Bouwmans2014}
\bibinfo{author}{Bouwmans, T.},  and \bibinfo{author}{Zahzah, E.-h.}
  (\textbf{\bibinfo{year}{2014}}). \enquote{\bibinfo{title}{Robust pca via
  principal component pursuit: A review for a comparative evaluation in video
  surveillance}} \bibinfo{journal}{Computer Vision and Image Understanding}
  \textbf{122}, \bibinfo{pages}{22--34},
  \dodoi{https://doi.org/10.1016/j.cviu.2013.11.009}.

\bibitem[{Brautaset \emph{et~al.}(2020)Brautaset, Waldeland, Johnsen, Malde,
  Eikvil, Salberg, and Handegard}]{Brautaset2020}
\bibinfo{author}{Brautaset, O.}, \bibinfo{author}{Waldeland, A.~U.},
  \bibinfo{author}{Johnsen, E.}, \bibinfo{author}{Malde, K.},
  \bibinfo{author}{Eikvil, L.}, \bibinfo{author}{Salberg, A.-B.},  and
  \bibinfo{author}{Handegard, N.~O.} (\textbf{\bibinfo{year}{2020}}).
  \enquote{\bibinfo{title}{{Acoustic classification in multifrequency
  echosounder data using deep convolutional neural networks}}}
  \bibinfo{journal}{ICES Journal of Marine Science} \textbf{77},
  \bibinfo{pages}{1391--1400}, \dodoi{doi.org/10.1093/icesjms/fsz235}.

\bibitem[{Brunet \emph{et~al.}(2004)Brunet, Tamayo, Golub, and
  Mesirov}]{Brunet2004}
\bibinfo{author}{Brunet, J.-P.}, \bibinfo{author}{Tamayo, P.},
  \bibinfo{author}{Golub, T.~R.},  and \bibinfo{author}{Mesirov, J.~P.}
  (\textbf{\bibinfo{year}{2004}}). \enquote{\bibinfo{title}{{Metagenes and
  molecular pattern discovery using matrix factorization.}}}
  \bibinfo{journal}{Proceedings of the National Academy of Sciences of the
  United States of America} \textbf{101}(12), \bibinfo{pages}{4164--9},
  \dodoi{10.1073/pnas.0308531101}.

\bibitem[{Cade and Benoit-Bird(2014)}]{Cade2014}
\bibinfo{author}{Cade, D.~E.},  and \bibinfo{author}{Benoit-Bird, K.~J.}
  (\textbf{\bibinfo{year}{2014}}). \enquote{\bibinfo{title}{{An automatic and
  quantitative approach to the detection and tracking of acoustic scattering
  layers}}} \bibinfo{journal}{Limnology and Oceanography: Methods}
  \textbf{12}(NOV), \bibinfo{pages}{742--756}, \dodoi{10.4319/lom.2014.12.742}.

\bibitem[{Candes \emph{et~al.}(2009)Candes, Li, Ma, and Wright}]{Candes2009}
\bibinfo{author}{Candes, E.~J.}, \bibinfo{author}{Li, X.}, \bibinfo{author}{Ma,
  Y.},  and \bibinfo{author}{Wright, J.} (\textbf{\bibinfo{year}{2009}}).
  \plainquote{\bibinfo{title}{{Robust Principal Component Analysis?}}}
  \bibinfo{howpublished}{http://arxiv.org/abs/0912.3599}.

\bibitem[{Chen \emph{et~al.}(2015)Chen, Zhang, Wu, Wang, Liu, and
  Lin}]{Chen2015}
\bibinfo{author}{Chen, Y.}, \bibinfo{author}{Zhang, H.}, \bibinfo{author}{Wu,
  J.}, \bibinfo{author}{Wang, X.}, \bibinfo{author}{Liu, R.},  and
  \bibinfo{author}{Lin, M.} (\textbf{\bibinfo{year}{2015}}).
  \enquote{\bibinfo{title}{{Modeling emerging, evolving and fading topics using
  dynamic soft orthogonal NMF with sparse representation}}} in
  \emph{\bibinfo{booktitle}{2015 IEEE International Conference on Data
  Mining}}, pp. \bibinfo{pages}{61--70},
  \dourl{http://ieeexplore.ieee.org/document/7373310/},
  \dodoi{10.1109/ICDM.2015.96}.

\bibitem[{Chen and Cichocki(2005)}]{Chen2005}
\bibinfo{author}{Chen, Z.},  and \bibinfo{author}{Cichocki, A.}
  (\textbf{\bibinfo{year}{2005}}). \enquote{\bibinfo{title}{{Nonnegative matrix
  factorization with temporal smoothness and/or spatial decorrelation
  constraints}}} in \emph{\bibinfo{booktitle}{Laboratory for Advanced Brain
  Signal Processing, RIKEN, Technical Report}},
  \dourl{http://citeseerx.ist.psu.edu/viewdoc/summary?doi=10.1.1.63.7094}.

\bibitem[{Chi(2018)}]{ShunChi}
\bibinfo{author}{Chi, S.} (\textbf{\bibinfo{year}{2018}}).
  \plainquote{\bibinfo{title}{Robust pca: Robust principle component analysis}}
  \bibinfo{howpublished}{https://github.com/ShunChi100/RobustPCA/}.

\bibitem[{Chu \emph{et~al.}(2019)Chu, Parker-Stetter, Hufnagle, Thomas,
  Getsiv-Clemons, Gauthier, and Stanley}]{Chu2019}
\bibinfo{author}{Chu, D.}, \bibinfo{author}{Parker-Stetter, S.},
  \bibinfo{author}{Hufnagle, L.~C.}, \bibinfo{author}{Thomas, R.},
  \bibinfo{author}{Getsiv-Clemons, J.}, \bibinfo{author}{Gauthier, S.},  and
  \bibinfo{author}{Stanley, C.} (\textbf{\bibinfo{year}{2019}}).
  \enquote{\bibinfo{title}{{2018 Unmanned Surface Vehicle (Saildrone) acoustic
  survey off the west coasts of the United States and Canada}}} in
  \emph{\bibinfo{booktitle}{OCEANS 2019 MTS/IEEE Seattle}},
  \dodoi{10.23919/OCEANS40490.2019.8962778}.

\bibitem[{Cichocki \emph{et~al.}(2009)Cichocki, Zdunek, Phan, and
  Amari}]{Cichocki2009}
\bibinfo{author}{Cichocki, A.}, \bibinfo{author}{Zdunek, R.},
  \bibinfo{author}{Phan, A.~H.},  and \bibinfo{author}{Amari, S.~I.}
  (\textbf{\bibinfo{year}{2009}}). \emph{\bibinfo{title}{{Nonnegative Matrix
  and Tensor Factorizations: Applications to Exploratory Multi-Way Data
  Analysis and Blind Source Separation}}} (\bibinfo{publisher}{John Wiley {\&}
  Sons, Inc.}), pp. \bibinfo{pages}{1--477}.

\bibitem[{{De Robertis} \emph{et~al.}(2018){De Robertis}, Levine, and
  Wilson}]{DeRobertis2018_mooring}
\bibinfo{author}{{De Robertis}, A.}, \bibinfo{author}{Levine, R.},  and
  \bibinfo{author}{Wilson, C.~D.} (\textbf{\bibinfo{year}{2018}}).
  \enquote{\bibinfo{title}{{Can a bottom-moored echo sounder array provide a
  survey-comparable index of abundance?}}} \bibinfo{journal}{Canadian Journal
  of Fisheries and Aquatic Sciences} \textbf{75}(4), \bibinfo{pages}{629--640},
  \dodoi{10.1139/cjfas-2017-0013}.

\bibitem[{De~Robertis \emph{et~al.}(2010)De~Robertis, McKelvey, and
  Ressler}]{DeRobertis2010_multifreq}
\bibinfo{author}{De~Robertis, A.}, \bibinfo{author}{McKelvey, D.~R.},  and
  \bibinfo{author}{Ressler, P.~H.} (\textbf{\bibinfo{year}{2010}}).
  \enquote{\bibinfo{title}{{Development and application of an empirical
  multifrequency method for backscatter classification}}}
  \bibinfo{journal}{Canadian Journal of Fisheries and Aquatic Sciences}
  \textbf{67}(9), \bibinfo{pages}{1459--1474}, \dodoi{10.1139/F10-075}.

\bibitem[{Demer \emph{et~al.}(2015)Demer, Berger, Bernasconi, Bethke, Boswell,
  Chu, Domokos, Dunford, Fassler, Gauthier, Hufnagle, Jech, Bouffant,
  Lebourges-Dhaaussy, Lurton, Macaulay, Perrot, Ryan, Parker-Stetter,
  Stienessen, Weber, and Williamson}]{Demer2015}
\bibinfo{author}{Demer, D.}, \bibinfo{author}{Berger, L.},
  \bibinfo{author}{Bernasconi, M.}, \bibinfo{author}{Bethke, E.},
  \bibinfo{author}{Boswell, K.~M.}, \bibinfo{author}{Chu, D.},
  \bibinfo{author}{Domokos, R.}, \bibinfo{author}{Dunford, A.},
  \bibinfo{author}{Fassler, S.}, \bibinfo{author}{Gauthier, S.},
  \bibinfo{author}{Hufnagle, L.}, \bibinfo{author}{Jech, J.},
  \bibinfo{author}{Bouffant, N.}, \bibinfo{author}{Lebourges-Dhaaussy, A.},
  \bibinfo{author}{Lurton, X.}, \bibinfo{author}{Macaulay, G.},
  \bibinfo{author}{Perrot, Y.}, \bibinfo{author}{Ryan, T.},
  \bibinfo{author}{Parker-Stetter, S.~L.}, \bibinfo{author}{Stienessen, S.},
  \bibinfo{author}{Weber, T.},  and \bibinfo{author}{Williamson, N.}
  (\textbf{\bibinfo{year}{2015}}). \enquote{\bibinfo{title}{{Calibration of
  acoustic instruments.}}} in \emph{\bibinfo{booktitle}{ICES Cooperative
  Research Report, No. 326}}, p. \bibinfo{pages}{133 pp},
  \dourl{http://ices.dk/sites/pub/Publication Reports/Cooperative Research
  Report (CRR)/crr326/CRR326.pdf}.

\bibitem[{Ding \emph{et~al.}(2008)Ding, Li, and Peng}]{Ding2008}
\bibinfo{author}{Ding, C.}, \bibinfo{author}{Li, T.},  and
  \bibinfo{author}{Peng, W.} (\textbf{\bibinfo{year}{2008}}).
  \enquote{\bibinfo{title}{{On the equivalence between Non-negative Matrix
  Factorization and probabilistic latent semantic indexing}}}
  \bibinfo{journal}{Computational Statistics and Data Analysis} \textbf{52}(8),
  \bibinfo{pages}{3913--3927}, \dodoi{10.1016/j.csda.2008.01.011}.

\bibitem[{Dunlop \emph{et~al.}(2018)Dunlop, Jarvis, Benoit-Bird, Waluk, Caress,
  Thomas, and Smith}]{Dunlop2018}
\bibinfo{author}{Dunlop, K.~M.}, \bibinfo{author}{Jarvis, T.},
  \bibinfo{author}{Benoit-Bird, K.~J.}, \bibinfo{author}{Waluk, C.~M.},
  \bibinfo{author}{Caress, D.~W.}, \bibinfo{author}{Thomas, H.},  and
  \bibinfo{author}{Smith, K.~L.} (\textbf{\bibinfo{year}{2018}}).
  \enquote{\bibinfo{title}{{Detection and characterisation of deep-sea
  benthopelagic animals from an autonomous underwater vehicle with a multibeam
  echosounder: A proof of concept and description of data-processing methods}}}
  \bibinfo{journal}{Deep-Sea Research Part I: Oceanographic Research Papers}
  \textbf{134}, \bibinfo{pages}{64--79}, \dodoi{10.1016/j.dsr.2018.01.006}.

\bibitem[{Fabregat \emph{et~al.}(2019)Fabregat, Pustelnik, Gon{\c{c}}alves, and
  Borgnat}]{Fabregat2019}
\bibinfo{author}{Fabregat, R.}, \bibinfo{author}{Pustelnik, N.},
  \bibinfo{author}{Gon{\c{c}}alves, P.},  and \bibinfo{author}{Borgnat, P.}
  (\textbf{\bibinfo{year}{2019}}). \plainquote{\bibinfo{title}{{Solving NMF
  with smoothness and sparsity constraints using PALM}}}
  \bibinfo{howpublished}{http://arxiv.org/abs/1910.14576}.

\bibitem[{Fallon \emph{et~al.}(2016)Fallon, Fielding, and
  Fernandes}]{Fallon2016}
\bibinfo{author}{Fallon, N.~G.}, \bibinfo{author}{Fielding, S.},  and
  \bibinfo{author}{Fernandes, P.~G.} (\textbf{\bibinfo{year}{2016}}).
  \enquote{\bibinfo{title}{{Classification of Southern Ocean krill and icefish
  echoes using random forests}}} \bibinfo{journal}{ICES Journal of Marine
  Science} \textbf{73}(8), \bibinfo{pages}{1998--2008},
  \dodoi{10.1093/icesjms/fsw057}.

\bibitem[{Fleischer \emph{et~al.}(2012)Fleischer, Ressler, Thomas, de~Blois,
  Hufnagle, and Chu}]{hake_survey_protocol_2012}
\bibinfo{author}{Fleischer, G.}, \bibinfo{author}{Ressler, P.},
  \bibinfo{author}{Thomas, R.}, \bibinfo{author}{de~Blois, S.},
  \bibinfo{author}{Hufnagle, L.},  and \bibinfo{author}{Chu, D.}
  (\textbf{\bibinfo{year}{2012}}). \enquote{\bibinfo{title}{{Pacific hake
  integrated acoustic and trawl survey methods}}} \bibinfo{type}{Technical
  Report}.

\bibitem[{Frigyesi and H{\"{o}}glund(2008)}]{Frigyesi2008}
\bibinfo{author}{Frigyesi, A.},  and \bibinfo{author}{H{\"{o}}glund, M.}
  (\textbf{\bibinfo{year}{2008}}). \enquote{\bibinfo{title}{{Non-negative
  matrix factorization for the analysis of complex gene expression data:
  identification of clinically relevant tumor subtypes.}}}
  \bibinfo{journal}{Cancer Informatics} \textbf{6}, \bibinfo{pages}{275--92},
  \dodoi{10.4137/cin.s606}.

\bibitem[{Greene \emph{et~al.}(2014)Greene, Meyer-Gutbrod, McGarry, Hufnagle,
  Chu, McClatchie, Packer, Jung, Acker, Dorn, and Pelkie}]{Greene2014}
\bibinfo{author}{Greene, C.~H.}, \bibinfo{author}{Meyer-Gutbrod, E.~L.},
  \bibinfo{author}{McGarry, L.~P.}, \bibinfo{author}{Hufnagle, L.~C.},
  \bibinfo{author}{Chu, D.}, \bibinfo{author}{McClatchie, S.},
  \bibinfo{author}{Packer, A.}, \bibinfo{author}{Jung, J.~B.},
  \bibinfo{author}{Acker, T.}, \bibinfo{author}{Dorn, H.},  and
  \bibinfo{author}{Pelkie, C.} (\textbf{\bibinfo{year}{2014}}).
  \enquote{\bibinfo{title}{{A wave glider approach to fisheries acoustics
  transforming how we monitor the nation's commercial fisheries in the 21st
  century}}} \bibinfo{journal}{Oceanography} \textbf{27}(4),
  \bibinfo{pages}{168--174}, \dodoi{10.5670/oceanog.2014.82}.

\bibitem[{Handegard \emph{et~al.}(2013)Handegard, Buisson, Brehmer, Chalmers,
  {De Robertis}, Huse, Kloser, Macaulay, Maury, Ressler, Stenseth, and
  God{\o}}]{Handegard2013}
\bibinfo{author}{Handegard, N.~O.}, \bibinfo{author}{Buisson, L.~D.},
  \bibinfo{author}{Brehmer, P.}, \bibinfo{author}{Chalmers, S.~J.},
  \bibinfo{author}{{De Robertis}, A.}, \bibinfo{author}{Huse, G.},
  \bibinfo{author}{Kloser, R.}, \bibinfo{author}{Macaulay, G.},
  \bibinfo{author}{Maury, O.}, \bibinfo{author}{Ressler, P.~H.},
  \bibinfo{author}{Stenseth, N.~C.},  and \bibinfo{author}{God{\o}, O.~R.}
  (\textbf{\bibinfo{year}{2013}}). \enquote{\bibinfo{title}{{Towards an
  acoustic-based coupled observation and modelling system for monitoring and
  predicting ecosystem dynamics of the open ocean}}} \bibinfo{journal}{Fish and
  Fisheries} \textbf{14}(4), \bibinfo{pages}{605--615},
  \dodoi{10.1111/j.1467-2979.2012.00480.x}.

\bibitem[{Haralabous and Georgakarakos(1996)}]{Haralabous1996}
\bibinfo{author}{Haralabous, J.},  and \bibinfo{author}{Georgakarakos, S.}
  (\textbf{\bibinfo{year}{1996}}). \enquote{\bibinfo{title}{{Artificial neural
  networks as a tool for species identification of fish schools}}}
  \bibinfo{journal}{ICES Journal of Marine Science} \textbf{53}(2),
  \bibinfo{pages}{173--180}, \dodoi{10.1006/jmsc.1996.0019}.

\bibitem[{Haris \emph{et~al.}(2018)Haris, Kloser, Ryan, and Malan}]{Haris2018}
\bibinfo{author}{Haris, K.}, \bibinfo{author}{Kloser, R.~J.},
  \bibinfo{author}{Ryan, T.~E.},  and \bibinfo{author}{Malan, J.}
  (\textbf{\bibinfo{year}{2018}}). \enquote{\bibinfo{title}{Deep-water
  calibration of echosounders used for biomass surveys and species
  identification}} \bibinfo{journal}{ICES Journal of Marine Science}
  \textbf{75}(3), \bibinfo{pages}{1117--1130}, \dodoi{10.1093/icesjms/fsx206}.

\bibitem[{Hays(2003)}]{Hays2003}
\bibinfo{author}{Hays, G.~C.} (\textbf{\bibinfo{year}{2003}}).
  \enquote{\bibinfo{title}{{A review of the adaptive significance and ecosystem
  consequences of zooplankton diel vertical migrations}}} in
  \emph{\bibinfo{booktitle}{Migrations and Dispersal of Marine Organisms}}
  (\bibinfo{publisher}{Springer Netherlands}), pp. \bibinfo{pages}{163--170},
  \dodoi{10.1007/978-94-017-2276-6_18}.

\bibitem[{Hutchins \emph{et~al.}(2008)Hutchins, Murphy, Singh, and
  Graber}]{Hutchins2008}
\bibinfo{author}{Hutchins, L.~N.}, \bibinfo{author}{Murphy, S.~M.},
  \bibinfo{author}{Singh, P.},  and \bibinfo{author}{Graber, J.~H.}
  (\textbf{\bibinfo{year}{2008}}). \enquote{\bibinfo{title}{{Position-dependent
  motif characterization using non-negative matrix factorization.}}}
  \bibinfo{journal}{Bioinformatics} \textbf{24}(23), \bibinfo{pages}{2684--90},
  \dodoi{10.1093/bioinformatics/btn526}.

\bibitem[{Jech and Michaels(2006)}]{Jech2006}
\bibinfo{author}{Jech, J.~M.},  and \bibinfo{author}{Michaels, W.~L.}
  (\textbf{\bibinfo{year}{2006}}). \enquote{\bibinfo{title}{{A multifrequency
  method to classify and evaluate fisheries acoustics data}}}
  \bibinfo{journal}{Canadian Journal of Fisheries and Aquatic Sciences}
  \textbf{63}(10), \bibinfo{pages}{2225--2235}, \dodoi{10.1139/F06-126}.

\bibitem[{Keister \emph{et~al.}(2009)Keister, Peterson, and
  Pierce}]{Keister2009}
\bibinfo{author}{Keister, J.~E.}, \bibinfo{author}{Peterson, W.~T.},  and
  \bibinfo{author}{Pierce, S.~D.} (\textbf{\bibinfo{year}{2009}}).
  \enquote{\bibinfo{title}{{Zooplankton distribution and cross-shelf transfer
  of carbon in an area of complex mesoscale circulation in the northern
  California Current}}} \bibinfo{journal}{Deep-Sea Research Part I:
  Oceanographic Research Papers} \textbf{56}(2), \bibinfo{pages}{212--231},
  \dodoi{10.1016/j.dsr.2008.09.004}.

\bibitem[{Kim and Tidor(2003)}]{Kim2003}
\bibinfo{author}{Kim, P.~M.},  and \bibinfo{author}{Tidor, B.}
  (\textbf{\bibinfo{year}{2003}}). \enquote{\bibinfo{title}{Subsystem
  identification through dimensionality reduction of large-scale gene
  expression data.}} \bibinfo{journal}{Genome research} \textbf{13}(7),
  \bibinfo{pages}{1706--18}, \dodoi{10.1101/gr.903503}.

\bibitem[{Klevjer \emph{et~al.}(2016)Klevjer, Irigoien, R{\o}stad, Fraile-Nuez,
  Ben{\'{i}}tez-Barrios, and Kaartvedt.}]{Klevjer2016}
\bibinfo{author}{Klevjer, T.~A.}, \bibinfo{author}{Irigoien, X.},
  \bibinfo{author}{R{\o}stad, A.}, \bibinfo{author}{Fraile-Nuez, E.},
  \bibinfo{author}{Ben{\'{i}}tez-Barrios, V.~M.},  and
  \bibinfo{author}{Kaartvedt., S.} (\textbf{\bibinfo{year}{2016}}).
  \enquote{\bibinfo{title}{{Large scale patterns in vertical distribution and
  behaviour of mesopelagic scattering layers}}} \bibinfo{journal}{Scientific
  Reports} \textbf{6}(1), \bibinfo{pages}{19873}, \dodoi{10.1038/srep19873}.

\bibitem[{{Koren} \emph{et~al.}(2009){Koren}, {Bell}, and
  {Volinsky}}]{Koren2009}
\bibinfo{author}{{Koren}, Y.}, \bibinfo{author}{{Bell}, R.},  and
  \bibinfo{author}{{Volinsky}, C.} (\textbf{\bibinfo{year}{2009}}).
  \enquote{\bibinfo{title}{Matrix factorization techniques for recommender
  systems}} \bibinfo{journal}{Computer} \textbf{42}(8),
  \bibinfo{pages}{30--37}.

\bibitem[{Korneliussen \emph{et~al.}(2016)Korneliussen, Heggelund, Macaulay,
  Patel, Johnsen, and Eliassen}]{Korneliussen2016}
\bibinfo{author}{Korneliussen, R.~J.}, \bibinfo{author}{Heggelund, Y.},
  \bibinfo{author}{Macaulay, G.~J.}, \bibinfo{author}{Patel, D.},
  \bibinfo{author}{Johnsen, E.},  and \bibinfo{author}{Eliassen, I.~K.}
  (\textbf{\bibinfo{year}{2016}}). \enquote{\bibinfo{title}{{Acoustic
  identification of marine species using a feature library}}}
  \bibinfo{journal}{Methods in Oceanography} \textbf{17},
  \bibinfo{pages}{187--205}, \dodoi{10.1016/j.mio.2016.09.002}.

\bibitem[{Lee and Seung(1999)}]{Lee1999_nmf}
\bibinfo{author}{Lee, D.~D.},  and \bibinfo{author}{Seung, H.~S.}
  (\textbf{\bibinfo{year}{1999}}). \enquote{\bibinfo{title}{{Learning the parts
  of objects by non-negative matrix factorization}}} \bibinfo{journal}{Nature}
  \textbf{401}(6755), \bibinfo{pages}{788--791}, \dodoi{10.1038/44565}.

\bibitem[{Lee \emph{et~al.}(2020)Lee, Nguyen, and Staneva}]{Lee2020_echopype}
\bibinfo{author}{Lee, W.-J.}, \bibinfo{author}{Nguyen, K.},  and
  \bibinfo{author}{Staneva, V.} (\textbf{\bibinfo{year}{2020}}).
  \plainquote{\bibinfo{title}{{Echopype: Enabling interoperability and
  scalability in ocean sonar data analysis (v0.4.0)}}}
  \bibinfo{howpublished}{https://zenodo.org/record/3907000},
  \dodoi{10.5281/ZENODO.3907000}.

\bibitem[{Lee and Staneva(2019)}]{Lee2019_tensor}
\bibinfo{author}{Lee, W.~J.},  and \bibinfo{author}{Staneva, V.}
  (\textbf{\bibinfo{year}{2019}}). \enquote{\bibinfo{title}{{Tensor
  decomposition of multi-frequency echosounder time series}}} in
  \emph{\bibinfo{booktitle}{OCEANS 2019 MTS/IEEE Seattle}},
  \bibinfo{publisher}{Institute of Electrical and Electronics Engineers Inc.},
  \dodoi{10.23919/OCEANS40490.2019.8962566}.

\bibitem[{Lehodey \emph{et~al.}(2010)Lehodey, Murtugudde, and
  Senina}]{Lehodey2010}
\bibinfo{author}{Lehodey, P.}, \bibinfo{author}{Murtugudde, R.},  and
  \bibinfo{author}{Senina, I.} (\textbf{\bibinfo{year}{2010}}).
  \enquote{\bibinfo{title}{{Bridging the gap from ocean models to population
  dynamics of large marine predators: A model of mid-trophic functional
  groups}}} \bibinfo{journal}{Progress in Oceanography} \textbf{84}(1-2),
  \bibinfo{pages}{69--84}, \dodoi{10.1016/J.POCEAN.2009.09.008}.

\bibitem[{Liutkus \emph{et~al.}(2015)Liutkus, Fitzgerald, and
  Badeau}]{Liutkus2015}
\bibinfo{author}{Liutkus, A.}, \bibinfo{author}{Fitzgerald, D.},  and
  \bibinfo{author}{Badeau, R.} (\textbf{\bibinfo{year}{2015}}).
  \enquote{\bibinfo{title}{{Cauchy nonnegative matrix factorization}}} in
  \emph{\bibinfo{booktitle}{2015 IEEE Workshop on Applications of Signal
  Processing to Audio and Acoustics (WASPAA)}},
  \dodoi{10.1109/WASPAA.2015.7336900}.

\bibitem[{Medwin and Clay(1998)}]{Medwin1998}
\bibinfo{author}{Medwin, H.},  and \bibinfo{author}{Clay, C.~S.}
  (\textbf{\bibinfo{year}{1998}}). \emph{\bibinfo{title}{{Fundamentals of
  acoustical oceanography}}} (\bibinfo{publisher}{Academic Press}), p.
  \bibinfo{pages}{712}.

\bibitem[{Mirkin(2011)}]{Mirkin2011}
\bibinfo{author}{Mirkin, B.} (\textbf{\bibinfo{year}{2011}}).
  \enquote{\bibinfo{title}{{Choosing the number of clusters}}}
  \bibinfo{journal}{WIREs Data Mining and Knowledge Discovery} \textbf{1}(3),
  \bibinfo{pages}{252--260}, \dodoi{10.1002/widm.15}.

\bibitem[{Mirzal(2014)}]{Mirzal2014}
\bibinfo{author}{Mirzal, A.} (\textbf{\bibinfo{year}{2014}}).
  \enquote{\bibinfo{title}{{Nonparametric Tikhonov regularized NMF and its
  application in cancer clustering}}} \bibinfo{journal}{IEEE/ACM Transactions
  on Computational Biology and Bioinformatics} \textbf{11}(6),
  \bibinfo{pages}{1208--1217}, \dodoi{10.1109/TCBB.2014.2328342}.

\bibitem[{Mohammadiha \emph{et~al.}(2015)Mohammadiha, Smaragdis, Panahandeh,
  and Doclo}]{Mohammadiha2015}
\bibinfo{author}{Mohammadiha, N.}, \bibinfo{author}{Smaragdis, P.},
  \bibinfo{author}{Panahandeh, G.},  and \bibinfo{author}{Doclo, S.}
  (\textbf{\bibinfo{year}{2015}}). \enquote{\bibinfo{title}{{A state-space
  approach to dynamic nonnegative matrix factorization}}}
  \bibinfo{journal}{IEEE Transactions on Signal Processing} \textbf{63}(4),
  \bibinfo{pages}{949--959}, \dodoi{10.1109/TSP.2014.2385655}.

\bibitem[{Moline \emph{et~al.}(2015)Moline, Benoit-Bird, O'Gorman, and
  Robbins}]{Moline2015}
\bibinfo{author}{Moline, M.~A.}, \bibinfo{author}{Benoit-Bird, K.},
  \bibinfo{author}{O'Gorman, D.},  and \bibinfo{author}{Robbins, I.~C.}
  (\textbf{\bibinfo{year}{2015}}). \enquote{\bibinfo{title}{{Integration of
  scientific echo sounders with an adaptable autonomous vehicle to extend our
  understanding of animals from the surface to the bathypelagic}}}
  \bibinfo{journal}{Journal of Atmospheric and Oceanic Technology}
  \textbf{32}(11), \bibinfo{pages}{2173--2186},
  \dodoi{10.1175/JTECH-D-15-0035.1}.

\bibitem[{{Ocean Observatories Initiative CE04OSBP}(2015)}]{OOI_CE04OSBP}
\bibinfo{author}{{Ocean Observatories Initiative CE04OSBP}}
  (\textbf{\bibinfo{year}{2015}}). \plainquote{\bibinfo{title}{{Oregon Offshore
  Cabled Benthic Experiment Package}}}
  \bibinfo{howpublished}{https://oceanobservatories.org/site/ce04osbp/}.

\bibitem[{{Ocean Observatories Initiative CE04OSPS}(2015)}]{OOI_CE04OSPS}
\bibinfo{author}{{Ocean Observatories Initiative CE04OSPS}}
  (\textbf{\bibinfo{year}{2015}}). \plainquote{\bibinfo{title}{{Oregon Offshore
  Cabled Shallow Profiler Mooring}}}
  \bibinfo{howpublished}{http://oceanobservatories.org/site/ce04osps/}.

\bibitem[{Oraintara \emph{et~al.}(2000)Oraintara, Karl, Castanon, and
  Nguyen}]{Oraintara2000}
\bibinfo{author}{Oraintara, S.}, \bibinfo{author}{Karl, W.},
  \bibinfo{author}{Castanon, D.},  and \bibinfo{author}{Nguyen, T.}
  (\textbf{\bibinfo{year}{2000}}). \enquote{\bibinfo{title}{A method for
  choosing the regularization parameter in generalized {Tikhonov} regularized
  linear inverse problems}} in \emph{\bibinfo{booktitle}{Proceedings 2000
  {International} {Conference} on {Image} {Processing}}}, Vol. 1, pp.
  \bibinfo{pages}{93--96 vol.1}, \dodoi{10.1109/ICIP.2000.900900},
  \bibinfo{note}{iSSN: 1522-4880}.

\bibitem[{Oseledets(2011)}]{Oseledets2011}
\bibinfo{author}{Oseledets, I.~V.} (\textbf{\bibinfo{year}{2011}}).
  \enquote{\bibinfo{title}{{Tensor-train decomposition}}}
  \bibinfo{journal}{SIAM Journal on Scientific Computing} \textbf{33}(5),
  \bibinfo{pages}{2295--2317}, \dodoi{10.1137/090752286}.

\bibitem[{Paatero and Tapper(1994)}]{Paatero1994}
\bibinfo{author}{Paatero, P.},  and \bibinfo{author}{Tapper, U.}
  (\textbf{\bibinfo{year}{1994}}). \enquote{\bibinfo{title}{{Positive matrix
  factorization: A non‐negative factor model with optimal utilization of
  error estimates of data values}}} \bibinfo{journal}{Environmetrics}
  \textbf{5}(2), \bibinfo{pages}{111--126}, \dodoi{10.1002/env.3170050203}.

\bibitem[{Parra \emph{et~al.}(2019)Parra, Greer, Book, Deary, Soto, Culpepper,
  Hernandez, and Miles}]{Parra2019}
\bibinfo{author}{Parra, S.~M.}, \bibinfo{author}{Greer, A.~T.},
  \bibinfo{author}{Book, J.~W.}, \bibinfo{author}{Deary, A.~L.},
  \bibinfo{author}{Soto, I.~M.}, \bibinfo{author}{Culpepper, C.},
  \bibinfo{author}{Hernandez, F.~J.},  and \bibinfo{author}{Miles, T.~N.}
  (\textbf{\bibinfo{year}{2019}}). \enquote{\bibinfo{title}{{Acoustic detection
  of zooplankton diel vertical migration behaviors on the northern Gulf of
  Mexico shelf}}} \bibinfo{journal}{Limnology and Oceanography} \textbf{64}(5),
  \bibinfo{pages}{2092--2113}, \dodoi{10.1002/lno.11171}.

\bibitem[{Pnevmatikakis \emph{et~al.}(2016)Pnevmatikakis, Soudry, Gao, Machado,
  Merel, Pfau, Reardon, Mu, Lacefield, Yang, Ahrens, Bruno, Jessell, Peterka,
  Yuste, and Paninski}]{Pnevmatikakis2016}
\bibinfo{author}{Pnevmatikakis, E.~A.}, \bibinfo{author}{Soudry, D.},
  \bibinfo{author}{Gao, Y.}, \bibinfo{author}{Machado, T.~A.},
  \bibinfo{author}{Merel, J.}, \bibinfo{author}{Pfau, D.},
  \bibinfo{author}{Reardon, T.}, \bibinfo{author}{Mu, Y.},
  \bibinfo{author}{Lacefield, C.}, \bibinfo{author}{Yang, W.},
  \bibinfo{author}{Ahrens, M.}, \bibinfo{author}{Bruno, R.},
  \bibinfo{author}{Jessell, T.~M.}, \bibinfo{author}{Peterka, D.~S.},
  \bibinfo{author}{Yuste, R.},  and \bibinfo{author}{Paninski, L.}
  (\textbf{\bibinfo{year}{2016}}). \enquote{\bibinfo{title}{{Simultaneous
  denoising, deconvolution, and demixing of calcium imaging data}}}
  \bibinfo{journal}{Neuron} \textbf{89}(2), \bibinfo{pages}{285--99},
  \dodoi{10.1016/j.neuron.2015.11.037}.

\bibitem[{Proud \emph{et~al.}(2015)Proud, Cox, Wotherspoon, and
  Brierley}]{Proud2015}
\bibinfo{author}{Proud, R.}, \bibinfo{author}{Cox, M.~J.},
  \bibinfo{author}{Wotherspoon, S.},  and \bibinfo{author}{Brierley, A.~S.}
  (\textbf{\bibinfo{year}{2015}}). \enquote{\bibinfo{title}{{A method for
  identifying Sound Scattering Layers and extracting key characteristics}}}
  \bibinfo{journal}{Methods in Ecology and Evolution} \textbf{6}(10),
  \bibinfo{pages}{1190--1198}, \dodoi{10.1111/2041-210X.12396}.

\bibitem[{Saha and Sindhwani(2012)}]{Saha2012}
\bibinfo{author}{Saha, A.},  and \bibinfo{author}{Sindhwani, V.}
  (\textbf{\bibinfo{year}{2012}}). \enquote{\bibinfo{title}{{Learning evolving
  and emerging topics in social media: A dynamic nmf approach with temporal
  regularization}}} in \emph{\bibinfo{booktitle}{ACM international
  conference}}, pp. \bibinfo{pages}{693--702}, \dodoi{10.1145/2124295.2124376}.

\bibitem[{Simmonds and MacLennan(2007)}]{Simmonds2005}
\bibinfo{author}{Simmonds, J.},  and \bibinfo{author}{MacLennan, D.}
  (\textbf{\bibinfo{year}{2007}}). \emph{\bibinfo{title}{{Fisheries acoustics:
  Theory and practice: Second edition}}} (\bibinfo{publisher}{Blackwell
  Science}), pp. \bibinfo{pages}{1--252}.

\bibitem[{Sokal and Rohlf(1962)}]{Sokal1962}
\bibinfo{author}{Sokal, R.~R.},  and \bibinfo{author}{Rohlf, F.~J.}
  (\textbf{\bibinfo{year}{1962}}). \enquote{\bibinfo{title}{The comparison of
  dendrograms by objective methods}} \bibinfo{journal}{TAXON} \textbf{11}(2),
  \bibinfo{pages}{33--40},
  \dourl{https://onlinelibrary.wiley.com/doi/abs/10.2307/1217208},
  \dodoi{10.2307/1217208}.

\bibitem[{Staneva and Lee(2020)}]{Staneva2020}
\bibinfo{author}{Staneva, V.},  and \bibinfo{author}{Lee, W.-J.}
  (\textbf{\bibinfo{year}{2020}}). \plainquote{\bibinfo{title}{time-series-nmf:
  Non-negative matrix factorization for time series (v0.1.0.dev0)}}
  \bibinfo{howpublished}{http://doi.org/10.5281/zenodo.3906891},
  \dodoi{10.5281/zenodo.3906891}.

\bibitem[{Suberg \emph{et~al.}(2014)Suberg, Wynn, Kooij, Fernand, Fielding,
  Guihen, Gillespie, Johnson, Gkikopoulou, Allan, Vrana, Miller, Smeed, and
  Jones}]{Suberg2014}
\bibinfo{author}{Suberg, L.}, \bibinfo{author}{Wynn, R.~B.},
  \bibinfo{author}{Kooij, J. V.~D.}, \bibinfo{author}{Fernand, L.},
  \bibinfo{author}{Fielding, S.}, \bibinfo{author}{Guihen, D.},
  \bibinfo{author}{Gillespie, D.}, \bibinfo{author}{Johnson, M.},
  \bibinfo{author}{Gkikopoulou, K.~C.}, \bibinfo{author}{Allan, I.~J.},
  \bibinfo{author}{Vrana, B.}, \bibinfo{author}{Miller, P.~I.},
  \bibinfo{author}{Smeed, D.},  and \bibinfo{author}{Jones, A.~R.}
  (\textbf{\bibinfo{year}{2014}}). \enquote{\bibinfo{title}{{Assessing the
  potential of autonomous submarine gliders for ecosystem monitoring across
  multiple trophic levels (plankton to cetaceans) and pollutants in shallow
  shelf seas}}} \bibinfo{journal}{Methods in Oceanography} \textbf{10},
  \bibinfo{pages}{70--89}, \dodoi{10.1016/j.mio.2014.06.002}.

\bibitem[{Turk and Pentland(1991)}]{Turk1991}
\bibinfo{author}{Turk, M.},  and \bibinfo{author}{Pentland, A.}
  (\textbf{\bibinfo{year}{1991}}). \enquote{\bibinfo{title}{{Face recognition
  using eigenfaces}}} in \emph{\bibinfo{booktitle}{Proceedings of the 1991 IEEE
  Computer Society Conference on Computer Vision and Pattern Recognition}}, pp.
  \bibinfo{pages}{586--591}, \dodoi{10.1109/CVPR.1991.139758}.

\bibitem[{Urmy \emph{et~al.}(2012)Urmy, Horne, and Barbee}]{Urmy2012}
\bibinfo{author}{Urmy, S.~S.}, \bibinfo{author}{Horne, J.~K.},  and
  \bibinfo{author}{Barbee, D.~H.} (\textbf{\bibinfo{year}{2012}}).
  \enquote{\bibinfo{title}{{Measuring the vertical distributional variability
  of pelagic fauna in Monterey Bay}}} \bibinfo{journal}{ICES Journal of Marine
  Science} \textbf{69}(2), \bibinfo{pages}{184--196},
  \dodoi{10.1093/icesjms/fsr205}.

\bibitem[{Ward(1963)}]{Ward1963}
\bibinfo{author}{Ward, Jr., J.~H.} (\textbf{\bibinfo{year}{1963}}).
  \enquote{\bibinfo{title}{Hierarchical grouping to optimize an objective
  function}} \bibinfo{journal}{Journal of the American Statistical Association}
  \textbf{58}(301), \bibinfo{pages}{236--244},
  \dodoi{10.1080/01621459.1963.10500845}.

\bibitem[{Woillez \emph{et~al.}(2007)Woillez, Poulard, Rivoirard, Petitgas, and
  Bez}]{Woillez2007}
\bibinfo{author}{Woillez, M.}, \bibinfo{author}{Poulard, J.~C.},
  \bibinfo{author}{Rivoirard, J.}, \bibinfo{author}{Petitgas, P.},  and
  \bibinfo{author}{Bez, N.} (\textbf{\bibinfo{year}{2007}}).
  \enquote{\bibinfo{title}{{Indices for capturing spatial patterns and their
  evolution in time, with application to European hake (\emph{Merluccius
  merluccius}) in the Bay of Biscay}}} \bibinfo{journal}{ICES Journal of Marine
  Science} \textbf{64}(3), \bibinfo{pages}{537--550},
  \dodoi{10.1093/icesjms/fsm025}.

\bibitem[{Woillez \emph{et~al.}(2012)Woillez, Ressler, Wilson, and
  Horne}]{Woillez2012}
\bibinfo{author}{Woillez, M.}, \bibinfo{author}{Ressler, P.~H.},
  \bibinfo{author}{Wilson, C.~D.},  and \bibinfo{author}{Horne, J.~K.}
  (\textbf{\bibinfo{year}{2012}}). \enquote{\bibinfo{title}{{Multifrequency
  species classification of acoustic-trawl survey data using semi-supervised
  learning with class discovery}}} \bibinfo{journal}{The Journal of the
  Acoustical Society of America} \textbf{131}(2),
  \bibinfo{pages}{EL184--EL190}, \dodoi{10.1121/1.3678685}.

\bibitem[{Wu \emph{et~al.}(2014)Wu, Zhou, Pierce, Barth, and Cowles}]{Wu2014}
\bibinfo{author}{Wu, D.}, \bibinfo{author}{Zhou, M.}, \bibinfo{author}{Pierce,
  S.}, \bibinfo{author}{Barth, J.},  and \bibinfo{author}{Cowles, T.}
  (\textbf{\bibinfo{year}{2014}}). \enquote{\bibinfo{title}{{Zooplankton
  distribution and transport in the California Current off Oregon}}}
  \bibinfo{journal}{Marine Ecology Progress Series} \textbf{508},
  \bibinfo{pages}{87--103}, \dodoi{10.3354/meps10835}.

\bibitem[{Yang \emph{et~al.}(2011)Yang, Zhang, Yuan, and Oja}]{Yang2011}
\bibinfo{author}{Yang, Z.}, \bibinfo{author}{Zhang, H.}, \bibinfo{author}{Yuan,
  Z.},  and \bibinfo{author}{Oja, E.} (\textbf{\bibinfo{year}{2011}}).
  \enquote{\bibinfo{title}{Kullback-leibler divergence for nonnegative matrix
  factorization}} in \emph{\bibinfo{booktitle}{Artificial Neural Networks and
  Machine Learning -- ICANN 2011}}, edited by \bibinfo{editor}{T.~Honkela},
  \bibinfo{editor}{W.~Duch}, \bibinfo{editor}{M.~Girolami}, and
  \bibinfo{editor}{S.~Kaski}, \bibinfo{publisher}{Springer Berlin Heidelberg},
  \bibinfo{address}{Berlin, Heidelberg}, pp. \bibinfo{pages}{250--257}.

\bibitem[{Zhang \emph{et~al.}(2006)Zhang, Wang, Ford, and Makedon}]{Zhang2006}
\bibinfo{author}{Zhang, S.}, \bibinfo{author}{Wang, W.}, \bibinfo{author}{Ford,
  J.},  and \bibinfo{author}{Makedon, F.} (\textbf{\bibinfo{year}{2006}}).
  \enquote{\bibinfo{title}{Learning from incomplete ratings using non-negative
  matrix factorization}} in \emph{\bibinfo{booktitle}{SDM}}.

\bibitem[{Zhou \emph{et~al.}(2010)Zhou, Li, Wright, Cand{\`e}s, and
  Ma}]{Zhou2010}
\bibinfo{author}{Zhou, Z.}, \bibinfo{author}{Li, X.}, \bibinfo{author}{Wright,
  J.}, \bibinfo{author}{Cand{\`e}s, E.},  and \bibinfo{author}{Ma, Y.}
  (\textbf{\bibinfo{year}{2010}}). \enquote{\bibinfo{title}{Stable principal
  component pursuit}} \bibinfo{journal}{2010 IEEE International Symposium on
  Information Theory} \bibinfo{pages}{1518--1522}.

\end{thebibliography}
\bibliographystyle{jasaauthyear2}

\newpage


\clearpage
\newpage

\section*{Supplementary material}

\renewcommand{\thefigure}{S\arabic{figure}}
\setcounter{figure}{0}

\setcounter{page}{1}

\begin{figure}[!htb]
	\centering
	\includegraphics[width=\textwidth]{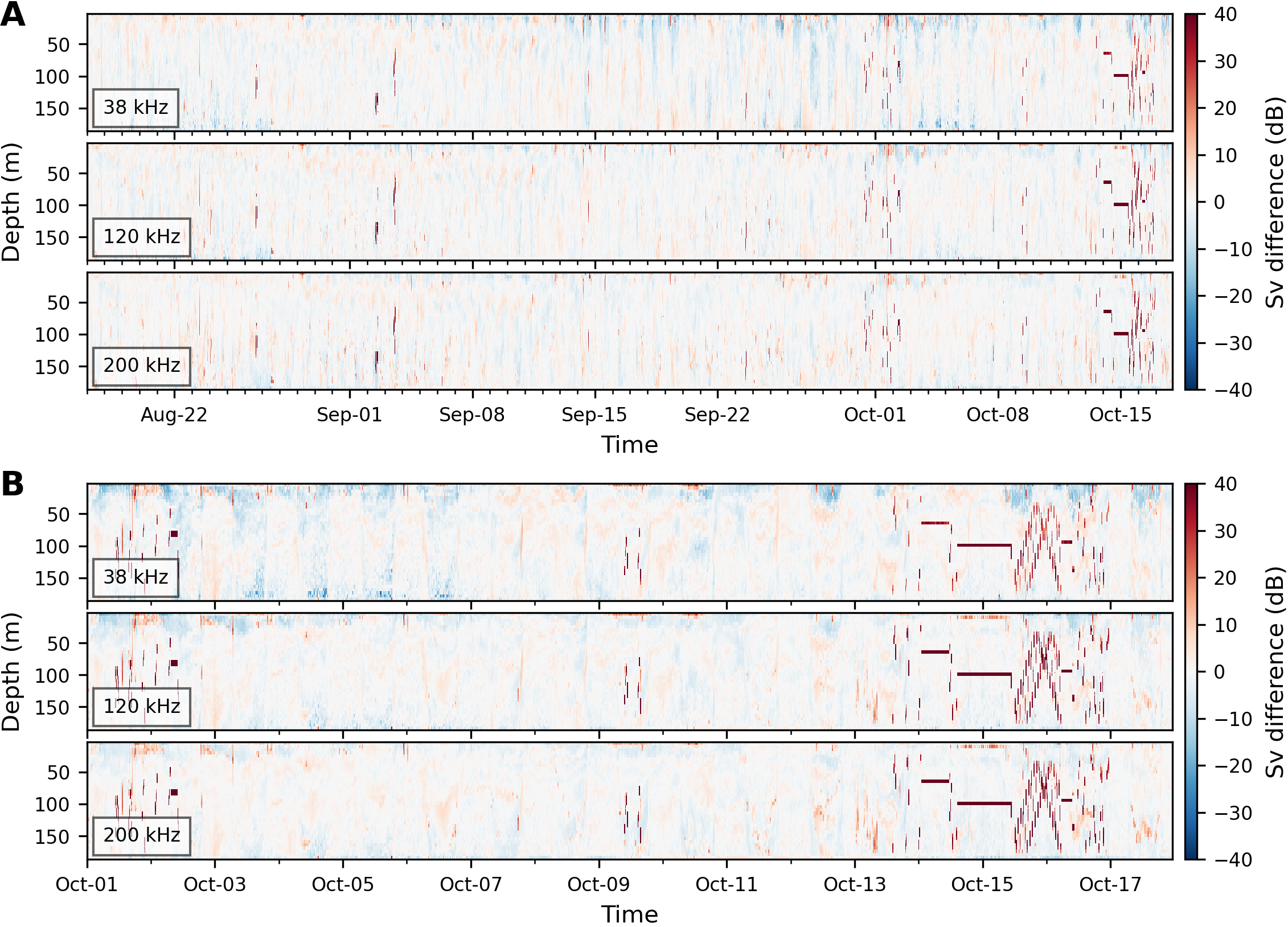}
	\caption{(A) Sparse component resulting from Principal Component Pursuit (PCP) analysis of the MVBS time series. The strong and irregular echo traces from the water column profiler are easily observed in mid October and also occurred sporadically earlier. (B) A zoomed-in section showing examples of other fine-grained echogram features also contained in the sparse component.}
	\label{fig:si_pcp_sparse}
\end{figure}

\begin{figure}[!htb]
	\centering
	\includegraphics[width=\textwidth]{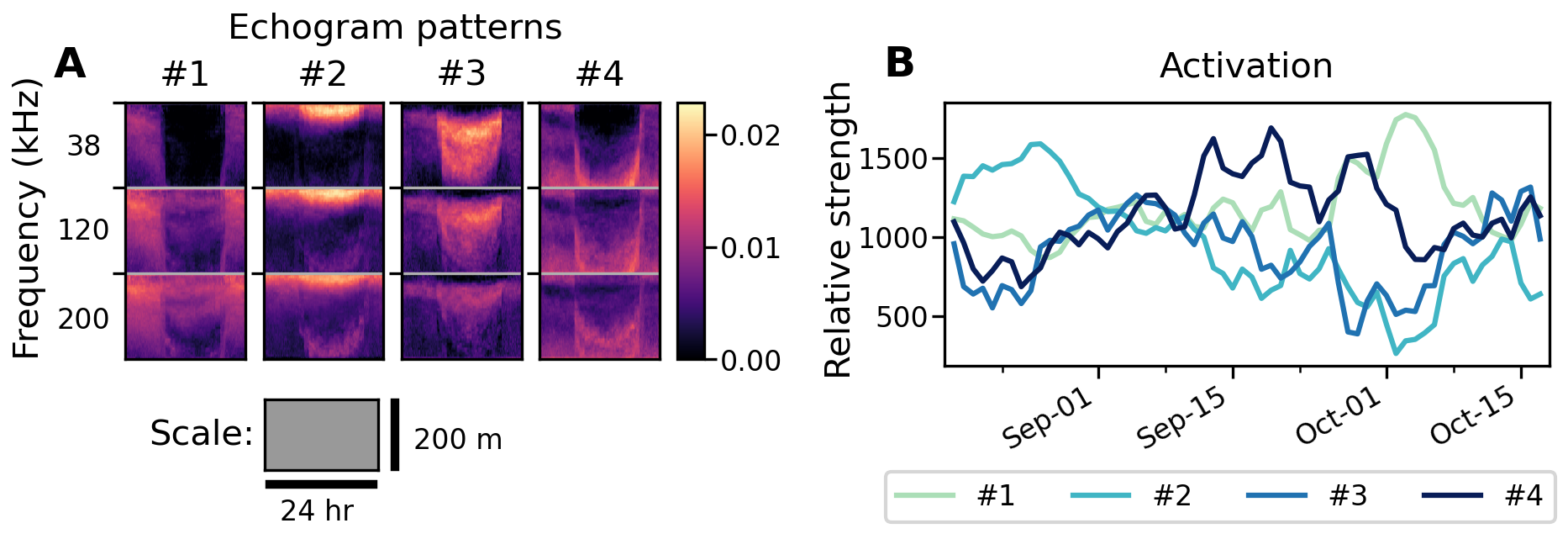}
	\caption{Decomposition results of tsNMF at rank=4. (A) Daily echogram patterns. (B) Activation sequences of the daily echogram patterns. All other details are identical to Fig.~\ref{fig:results_decomp}. Compared to the decomposition results at rank=3, we observe that Pattern \#2 and \#3 remain largely stable, whereas Pattern \#1 at rank=3 together with additional finer echogram features are decomposed into new patterns \#1 and \#4, representing a finer dissection of the DVM at rank=4.}
	\label{fig:si_rank_4}
\end{figure}

\begin{figure}[!htb]
	\centering
	\includegraphics[width=\textwidth]{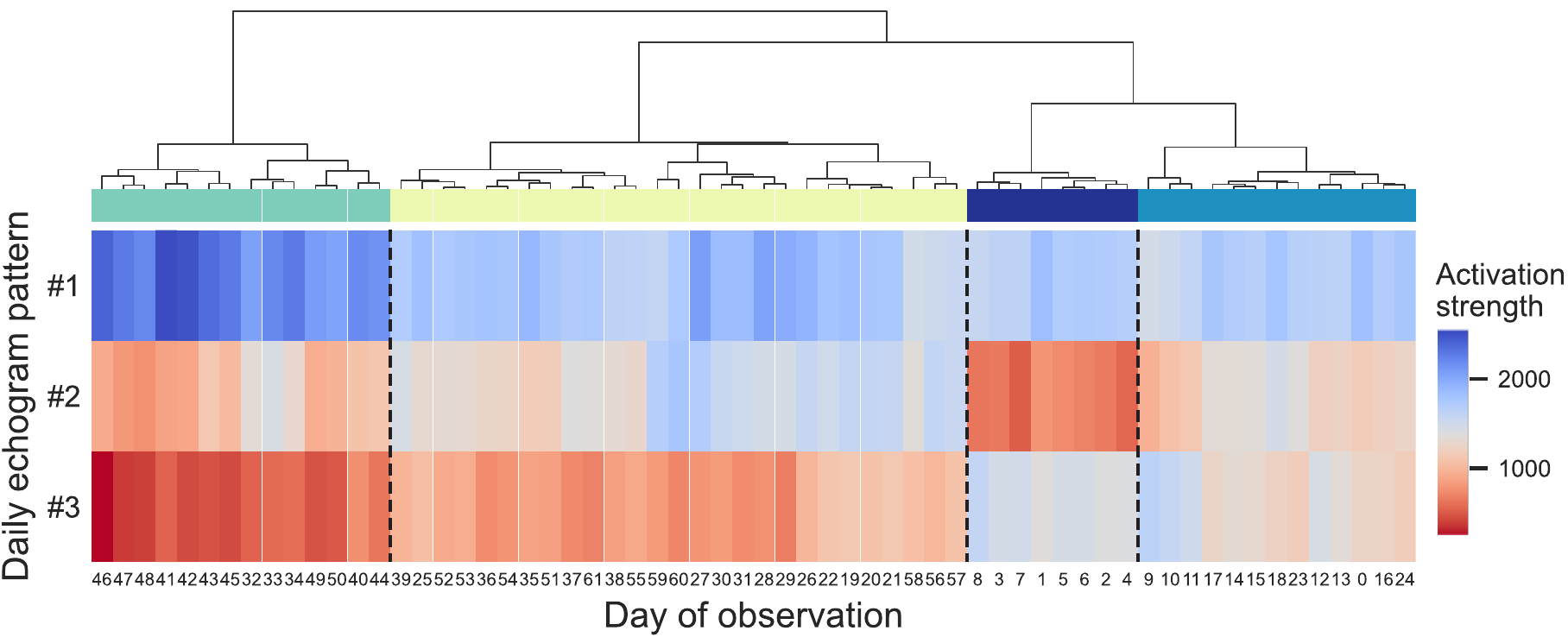}
	\caption{Hierarchical agglomerative clustering of the activation strengths of daily echogram patterns. The clustering is computed based on the Euclidean distance between the daily activations and uses the Ward's minimum variance method which minimizes the within-cluster variance when clusters are merged. In a dendrogram, the height at which any two objects are first joined indicates the distance between their corresponding clusters. We cut the dendrogram at a level where there is substantial separation between the clusters and obtain four clusters. The clusters are color-coded in the same way as in Fig.~\ref{fig:distance_matrix}B, with the vertical dashed lines on the heatmap delineates their boundaries. Changes of the cluster membership coincides with changes of ocean current direction at multiple time points throughout the observation period (Fig.~\ref{fig:distance_matrix}B-C).}
	\label{fig:si_clustering_dendrogram}
\end{figure}

\begin{figure}[!htb]
	\centering
	\includegraphics[width=0.5\textwidth]{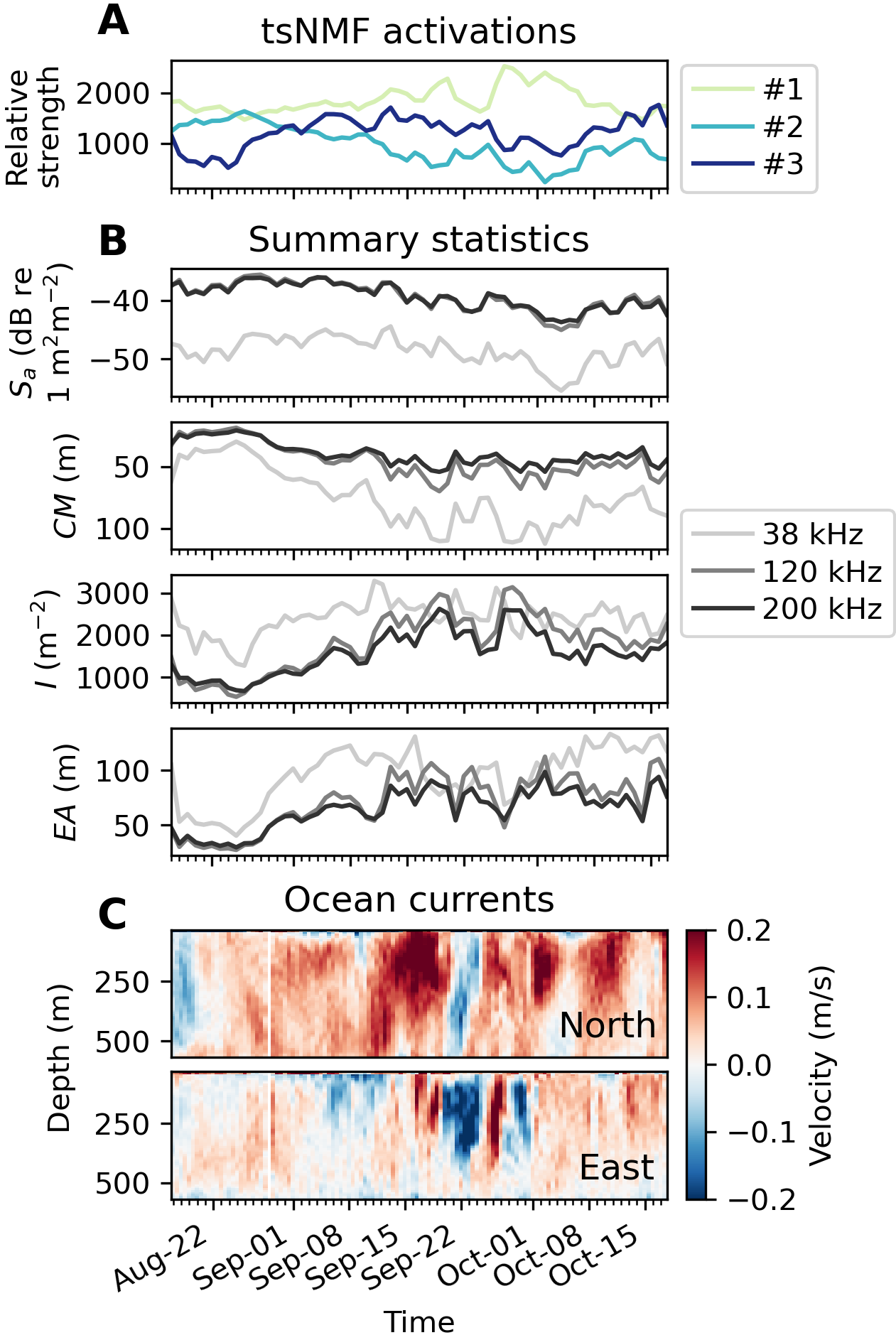}
	\caption{(A) tsNMF activation sequences as shown in Fig.~\ref{fig:results_decomp}. (B) Time series of four echogram summary statistics, including area backscattering strength ($S_a$), center of mass ($CM$), inertia ($I$) and equivalent area ($EA$), which provide proxies for the abundance of the observed animal aggregation and the weighted mean depth, dispersion and evenness of their vertical distribution, respectively \cite{Urmy2012, Woillez2007}. The time series are daily averages of the statistics computed using the low-rank MVBS data to avoid contamination by the profiler echoes. (C) Ocean currents as shown in Fig.~\ref{fig:distance_matrix}C. Note that the statistics could each be associated with only a subset of changes in ocean currents, but it is difficult to infer the actual echogram structure based solely on these statistics. In addition, they are calculated separately for each sonar frequency, whereas features from all three frequencies are inherently incorporated in each tsNMF component due to the decomposition formulation.}
	\label{fig:si_summary_stat}
\end{figure}

\begin{figure}[!htb]
	\centering
	\includegraphics[width=\textwidth]{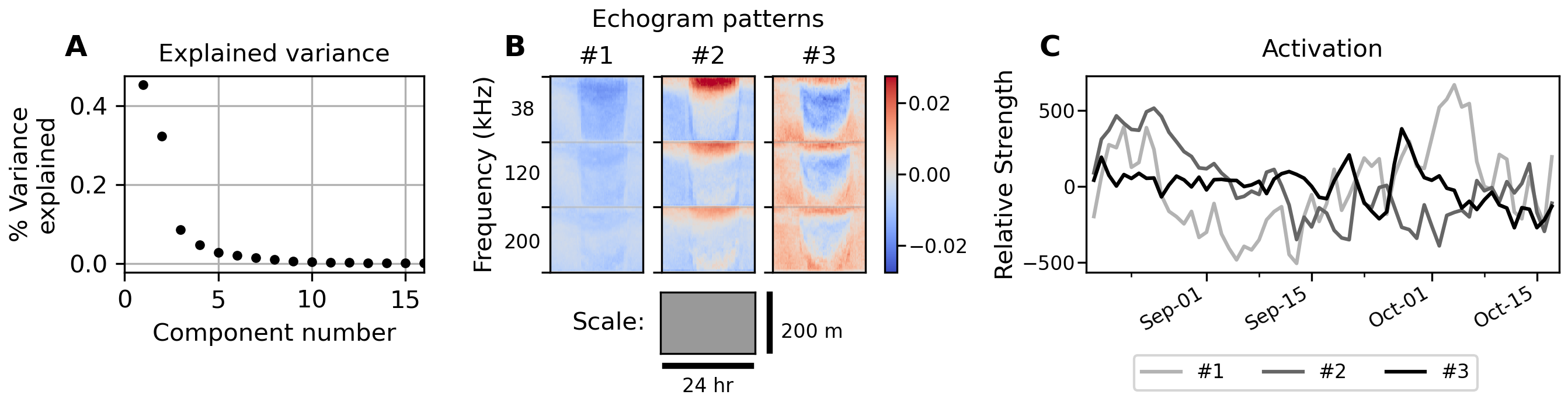}
	\caption{Decomposition results of PCA. (A) Percentage of variance explained by each principal component. (B) Daily echogram patterns of the first three principal components. (C) Activation coefficients of the first three principal components. Note the colormap for daily echogram patterns is changed to a bi-directional diverging scheme to highlight the fact that PCA components contain both positive and negative entries, instead of the solely nonnegative entries from tsNMF. While the first two PCA components resemble the first two from tsNMF (Fig.~\ref{fig:results_decomp}), their contributions to the observed MVBS echogram are much more difficult to interpret due to the intermingled positive and negative entries in both the patterns and the activation sequences.}
	\label{fig:si_pca}
\end{figure}

\begin{figure}[!htb]
	\centering
	\includegraphics[width=\textwidth]{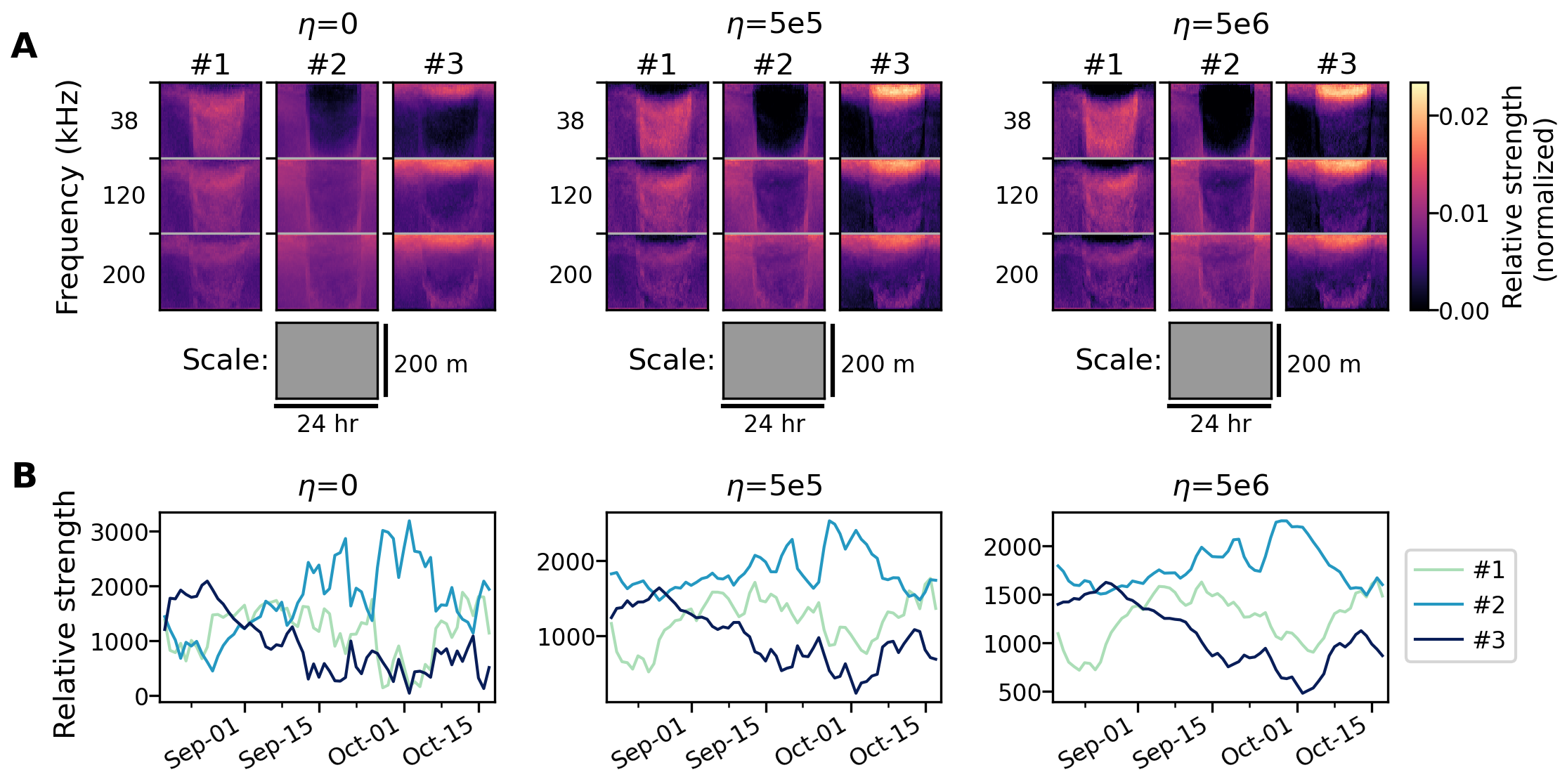}
	\caption{Effects of adjusting the smoothness parameter $\eta$ in tsNMF. (A) Daily echogram patterns. (B) Activation sequences of the daily echogram patterns. All other details are identical to Fig.~\ref{fig:results_decomp}. Note that while the main structures of the daily echogram patterns do not change dramatically, the smoother activation sequences obtained with higher $\eta$ are easier to interpret for longer time series.}
	\label{fig:si_sm_effects}
\end{figure}

\end{document}